\pdfoutput=1
\documentclass[a4paper,12pt]{article}
\usepackage{ifpdf}
\ifpdf
\usepackage[pdftex]{graphicx}
\usepackage[usenames,dvipsnames]{color}
\else
\usepackage{graphicx}
\usepackage[usenames,dvips]{color}
\fi
\usepackage{amsmath}
\usepackage{amssymb}
\usepackage{amsfonts,amsbsy,amscd,amsopn,amstext,amsxtra,latexsym,}
\usepackage[colorlinks=true]{hyperref}   
\usepackage{caption}
\usepackage{paper}

\numberwithin{equation}{section}

\begin{document}

\def\TimeDefectTraditional{1.00}
\def\TimeDefectParallel{0.58}
\def\TimeDefectStandard{0.57}
\def\TimeDerivsExact{1.11}
\def\TimeDerivsApprx{6.03}

\title{Fast algorithms for computing defects and their derivatives in the Regge calculus.}
\author{%
Leo Brewin\\[10pt]%
School of Mathematical Sciences\\%
Monash University, 3800\\%
Australia}
\date{08-Jan-2010}
\date{20-Jun-2010}
\date{10-Oct-2010}
\date{24-Oct-2010}
\date{29-Oct-2010}
\date{09-Nov-2010}
\reference{Preprint}

\maketitle

\begin{abstract}
\noindent
Any practical attempt to solve the Regge equations, these being a large system of non-linear
algebraic equations, will almost certainly employ a Newton-Raphson like scheme. In such cases
it is essential that efficient algorithms be used when computing the defect angles and their
derivatives with respect to the leg-lengths. The purpose of this paper is to present details
of such an algorithm.
\end{abstract}

\section{The Regge calculus}
\label{sec:ReggeCalc}

In its pure form (there are variations) the Regge calculus is a theory of gravity in which
the spacetime is built from a (possibly infinite) collection of non-overlapping flat
4-simplices (plus further low level constraints that ensure that the manifold remains
4-dimensional everywhere). This construction provides a clear distinction between the
topological and metric properties of such spacetimes. The topology is encoded in the way
the 4-simplices are tied together while the metric is expressed as an assignment of
lengths to each leg of the simplicial lattice.

The Regge action on a simplicial lattice is defined by
\begin{equation}
I_R = \sum_{\sigma_2} 2\left(\theta A\right)_{\sigma_2}
\label{eqn:ReggeAction}
\end{equation}
where $\theta_{\sigma_2}$ is the defect on a typical triangle $\sigma_2$, $A_{\sigma_2}$ is
the area of the triangle and the sum includes every triangle in the simplicial lattice. This
action is a function of the leg lengths $L^2(\sigma_1)$ and its extremization with respect to
a typical leg length leads to
\begin{equation}
0 = \sum_{\sigma_2(\sigma_1)} \left(\theta \pderiv(A,L^2)\right)_{\sigma_2}
\label{eqn:ReggeFieldEqtn}
\end{equation}
where the sum includes each triangle $\sigma_2$ that contains the leg $\sigma_1$. This set of
equations, one per leg, are known as the Regge field equations for the lattice. Constructing a
Regge simplicial lattice will require frequent computations of the defects as well as frequent
solutions of the above Regge equations, a large system of non-linear algebraic equations.
These are both nontrivial tasks.

Most articles on the Regge Calculus are concerned with formal issues, such as the nature of
its convergence to the continuum %
\cite{cheeger:1984-01,%
feinberg:1984-01,%
friedberg:1984-01,%
brewin:2000-01,%
miller-ma:1995-01}, 
its use as a tool in quantum gravity %
\cite{williams:2009-01,
rocek:1984-01}, %
the relationship of the Regge equations to the ADM equations %
\cite{piran:1986-01,friedman:1986-01} 
and so on (for further reading see %
\cite{williams:1992-01,
gentle:2002-01}).

When people turn to numerical computations in the Regge calculus they are usually concerned
with matters of existence (that solutions exist), accuracy (that the solutions compare well
with the continuum spacetime) and convergence (that a sequence of solutions converge to the
expected continuum spacetime). These questions are usually posed on sufficiently simple
lattices that the issue of computational efficiency is not a major concern. But in the end, if
the Regge calculus is to be of any use in numerical relativity, it must be at least
competitive with contemporary methods (\eg finite difference and spectral codes) in cases such
as the merger of binary black holes. Thus issues of computational efficiency will be of
paramount importance. It will not be sufficient to show that the Regge equations can be
solved. What will be more important is that the solutions can be found with minimal
computational effort to allow practical long-time evolutions to be undertaken.

To the best of the author's knowledge, there are no papers that pose the specific question
\emph{What is the best way to compute the defects?}. By \emph{best} we mean minimal
computational effort, that the cpu time required to compute each defect be as short as
possible. The main purpose of this article is to address that question. We do not
claim that our algorithms are optimal but rather that they are a significant improvement on
contemporary methods. In particular we will present details of an algorithm for computing
the derivatives of the defects at virtually no extra cost above that required to compute the
defects.

This paper is organised as follows. In section (\ref{sec:StdFrame}) we introduce various
elements such as a coordinate frame (the standard frame) and various vectors (\eg normal
vectors) that will be used later in section (\ref{sec:Defects}) to compute the defects. In
section (\ref{sec:Angles}) we provide a careful discussion regarding angles in a Lorentzian
spacetime. Explicit equations for the defects are given in section (\ref{sec:Defects}), and
their derivatives in section (\ref{sec:Derivs}). Finally, in section (\ref{sec:Timing}), we
present some simple timings for our algorithms against traditional methods.

Throughout this paper we will refer to simplices in two distinct ways. On occasions it will
be useful to identify the vertices that comprise a 4-simplex. In which case we will write
$(ijklm)$ for the 4-simplex built from the vertices $(i)$, $(j)$, $(k)$, $(l)$ and $(m)$. On
other occasions the vertices will be of lesser importance and so we will write $\sigma_4$ for
a typical $4-$simplex. Similar notation will be employed for other $n-$simplices.

We will adopt the following rules for coordinates indices, tetrad indices and vertex labels.
We will use Greek letters exclusively for coordinate indices (\eg for expressions valid
only in a specific coordinate frame, in particular the standard frame as defined in the
following section). However Roman letters will be used in two distinct ways and thus we choose
to divide the Roman alphabet into two groups. The first group, $a,b,c\cdots h$, will be used
as tetrad indices while the second group, $i,j,k\cdots$ will be used as vertex labels (\ie to
distinguish one vertex from another).

\section{The standard frame}
\label{sec:StdFrame}

It is rather easy to be seduced by the geometric elegance of the Regge calculus into
thinking that to maintain the purity of the formulation one must not only express but also
perform all computations in purely geometric form. As always, romance gives way to reality
and most people find it far easier to adopt local coordinates to simplify the
computations. The final results are always expressed solely in terms of the scalar data
(leg-lengths, angles etc.) and thus are independent of the chosen coordinate frame.

The frame which we are about to introduce, which we will refer to as the \emph{standard
frame}, is a trivial generalisation (from 3 to 4 dimensions) of a frame commonly used in
finite element computations on unstructured tetrahedral meshes. This frame has been
extensively used by others in the field (see for example \cite{piran:1986-01,gentle:2000-01}).

Consider a typical 4-simplex such as $(01234)$. For the standard frame, set the origin of the
coordinates at the vertex $(0)$, align the coordinate axes with the four edges and set
$e_a:=(0a), a=1,2,3,4$ as coordinate basis vectors. Now consider any point $x$ within this
4-simplex. Since the metric inside the 4-simplex is chosen to be flat we can uniquely express
the vector $(0x)$ as a linear combination of the $e_a$, namely
\begin{equation}
(0x) = x^\mu e_\mu
\label{eqn:StandardCoords}
\end{equation}
We take the $x^\mu$ to be the \emph{standard coordinates} of the point $x$. Note that these
coordinates are similar but distinct from another popular choice known as barycentric
coordinates (see \cite{friedman:1986-01,sorkin:1975-02,kheyfets:1989-01}).

Notice if the point $x$ lies \emph{on or inside} the 4-simplex then the
standard coordinates $x^\mu$ are subject to the constraints
\begin{equation*}
0\le x^\mu \le 1\qquad\text{and }\qquad 0\le x^1+x^2+x^3+x^4 \le 1
\end{equation*}
If the point $x$ happens to lie on one of the five faces of $(01234)$ then its standard
coordinates $x^\mu$ will be subject to the constraints listed in Table
(\ref{tbl:SimplexFaces})
\bgroup
\def\H{\vrule height 14pt depth  7pt width 0pt} 
\def\m{\vrule height  0pt depth 10pt width 0pt} 
\def\M{\vrule height 15pt depth 10pt width 0pt}
\begin{table}[ht]
\begin{center}
\begin{tabular}{ccccccc}
\hline
\H&Index&&Face&&Constraints&\\
\hline
\M&1&&$(0123)$&&$x^4 = 0$\\
\m&2&&$(0124)$&&$x^3 = 0$\\
\m&3&&$(0134)$&&$x^2 = 0$\\
\m&4&&$(0234)$&&$x^1 = 0$\\
\m&5&&$(1234)$&&$x^1+x^2+x^3+x^4 = 1$\\
\hline
\end{tabular}
\end{center}
\caption{The equations that describe the five faces of the 4-simplex $(01234)$ in the standard frame.}
\label{tbl:SimplexFaces}
\end{table}
\egroup

The corresponding constraints for the lower order simplices (\eg $(023)$) are given by
suitable combinations of the entries in that table (\eg for points on the face $(023)$ we take
the constraints for $(0123)$ and $(0234)$, \ie $x^1=0$ and $x^4=0$). Finally, we note that the
coordinates of the five vertices are simply given by
\begin{equation}
x_i^\mu = 
\begin{cases}
0&i = 0\\
\delta_i^\mu& i=1,2,3,4
\end{cases}
\label{eqn:StdVertexCoords}
\end{equation}

The flat metric for this 4-simplex can be constructed from the leg-lengths by choosing a
constant symmetric $4 \times 4$ matrix $g_{\mu\nu}$ such that
\begin{equation*}
\Lsqij = g_{\mu\nu}\Dxij^\mu\Dxij^\nu\qquad\qquad i,j=0,1,2,3,4
\end{equation*}
where $\Dxij^\mu := x^\mu_j -x^\mu_i$. This leads to a $10 \times 10$ system of equations for
the $g_{\mu\nu}$ and the solution is easily seen to be
\begin{equation}
g_{ij} =\half\left(\Lsq_{0i}+\Lsq_{0j}-\Lsq_{ij}\right)\qquad i,j=1,2,3,4
\label{eqn:StandardMetric}
\end{equation}
provided we take $L_{ij}=0$ when $i=j$.

This completes the definition of the standard frame for the chosen 4-simplex. A similar
construction can be applied to any other 4-simplex and it should be clear that the standard
frames of any pair of adjacent 4-simplices will be related by a linear transformation. We will
make use of this fact later in section (\ref{sec:Defects}) when we discuss the computation of
defect angles.

\subsection{Volumes}
\label{sec:Volumes}

The 4-volume $V_4$ of a typical 4-simplex is given by
\bgroup
\def\v{\vrule height9pt depth0pt width0pt}
\begin{align*}
V_4 &=\iiiint\limits_{\substack{\v0<x^1,x^2,x^3,x^4<1\\[2pt]
    0<x^1+x^2+x^3+x^4<1}}\>\sqrt{\abs{g}}\>dx^1dx^2dx^3dx^4\\
    &=\sqrt{\abs{g}}\>\iiiint\limits_{\substack{\v0<x^1,x^2,x^3,x^4<1\\[2pt]
    0<x^1+x^2+x^3+x^4<1}}\>dx^1dx^2dx^3dx^4\\
    &=\frac{1}{4!}\sqrt{\abs{g}}
\end{align*}
\egroup
where $g=\det(g_{\mu\nu})$ in the standard frame. Similar results can be derived for any
$n-$simplex. If we denote the $n-$volume (or measure) of the $n-$simplex by $V_n$ then
\begin{equation}
V_n = \frac{1}{n!} \sqrt{\abs{g}}
\label{eqn:StandardVolume}
\end{equation}
where $g=\det(g_{\mu\nu})$ in the standard frame for the $n-$simplex.

Another popular expression for the volume of a $n-$simplex is
\begin{equation*}
V_n = \frac{2^{-n/2}}{n!}\sqrt{\abs{W_n}}
\end{equation*}
where $W_n$ is the $(n+2)\times(n+2)$ Cayley-Menger determinant
\begin{equation*}
W_n =
\begin{vmatrix}
0&1&1&1&1&\dots&1\\[5pt]
1&0&L^2_{01}&L^2_{02}&L^2_{03}&\dots&L^2_{0n}\\[5pt]
1&L^2_{10}&0&L^2_{12}&L^2_{13}&\dots&L^2_{1n}\\[5pt]
1&L^2_{20}&L^2_{21}&0&L^2_{23}&\dots&L^2_{2n}\\[5pt]
1&L^2_{30}&L^2_{31}&L^2_{32}&0&\dots&L^2_{3n}\\
\hdotsfor{7}\\
1&L^2_{n0}&L^2_{n1}&L^2_{n3}&L^2_{n4}&\dots&0
\end{vmatrix}
\end{equation*}
It is a trivial exercise in linear algebra to show the equivalence of this pair of
equations for $V_n$.

There is also a useful relationship between $V_4$ and $V_3$ for each of the five faces of the
4-simplex. Let $(01234)$ be a typical 4-simplex and choose one of its five faces, say $(0123)$
as a base. Let $h$ be the height of $(01234)$ above the base $(0123)$. Then a simple extension
of \emph{base times height} theorem to higher dimensions gives
\begin{equation*}
V_4 = \frac{1}{4} h V_3
\end{equation*}
The height $h>0$ can be computed by taking the scalar projection of the edge vector (in this
case $(04)$) with the inward pointing unit normal to the face $(0123)$. If we denote the edge
and normal vectors by $e^a$ and $n^a$ respectively then we obtain
\begin{equation}
V_4 = \frac{1}{4}\abs{e^a n_a} V_3
\label{eqn:FourVol}
\end{equation}

\subsection{Normal vectors}
\label{sec:Normals}

Each 4-simplex is covered by 5 faces, each of which is a 3-simplex. How do we construct the
outward pointing unit normal to each face? Recall that if a plane is described by the equation
$n_\mu x^\mu =$ constant then its normal vector is parallel to $n^\mu$. Thus, by inspection of
table (\ref{tbl:SimplexFaces}) we find that components of the five normal vectors are as given
in table (\ref{tbl:Normals}).
\bgroup
\begin{table}[ht]
\def\H{\vrule height 14pt depth  7pt width 0pt} 
\def\m{\vrule height  0pt depth 10pt width 0pt} 
\def\M{\vrule height 15pt depth 10pt width 0pt}
\begin{center}
\begin{tabular}{ccccccc}
\hline
\H&Index&&Face&&Normal&\\
\hline
\M&1&&$(0123)$&&$n_{1\mu} = n_1\delta^4_\mu$\\
\m&2&&$(0124)$&&$n_{2\mu} = n_2\delta^3_\mu$\\
\m&3&&$(0134)$&&$n_{3\mu} = n_3\delta^2_\mu$\\
\m&4&&$(0234)$&&$n_{4\mu} = n_4\delta^1_\mu$\\
\m&5&&$(1234)$&&$n_{5\mu} = 
       n_5\left(\delta^1_\mu+\delta^2_\mu+\delta^3_\mu+\delta^4_\mu\right)$\\
\hline
\end{tabular}
\end{center}
\caption{%
Normal vectors to the five faces of $(01234)$. The $n_i$ are scale factors that ensure that
each vector is an outward pointing unit-vector.}
\label{tbl:Normals}
\end{table}
\egroup

Our next task is to compute each of the $n_i$ so that each $n_i^\mu$ is an outward
pointing unit vector.

If $n_i^\mu$ is a unit vector than we must have $\eps(n_i) = g_{\mu\nu} n_i^\mu n_i^\nu$ 
where $\eps(n_i) = \pm 1$ according to the signature of $n_i^\mu$. This leads to
\begin{spreadlines}{5pt}
\begin{gather*}
\eps(n_1) = n_1^2 g^{44} \qquad 
\eps(n_2) = n_2^2 g^{33} \\ 
\eps(n_3) = n_3^2 g^{22} \qquad
\eps(n_4) = n_4^2 g^{11} \qquad
\eps(n_5) = n_5^2 \sum_{\mu,\nu=1}^4\> g^{\mu\nu}
\end{gather*}
\end{spreadlines}

Since $n_i^2>0$ we see that that each $\eps(n_i)$ can be computed according to
\begin{spreadlines}{12pt}
\begin{gather*}
\eps(n_1) = \sign{g^{44}} \qquad 
\eps(n_2) = \sign{g^{33}} \\ 
\eps(n_3) = \sign{g^{22}} \qquad
\eps(n_4) = \sign{g^{11}} \qquad
\eps(n_5) = \sign{S}
\end{gather*}
\end{spreadlines}
where $S=\sum_{\mu,\nu=1}^4\> g^{\mu\nu}$ and the function $\sign{x}$ is defined by
\begin{equation*}
\sign{x} = 
\begin{cases}
+1&\text{for }x>0\\
-1&\text{for }x<0\\
0&\text{otherwise}
\end{cases}
\end{equation*}
All that remains is to choose an orientation for each $n_i^\mu$.

Consider for the moment the two sides of the face $(0123)$. Notice that the leg $(04)$ and
the 4-simplex $(01234)$ lie on the same side of this face. Thus if we want $n_1^\mu$ to be
outward pointing we simply demand that
\begin{equation*}
\eps(n_1) n_{1\mu} (04)^\mu < 0
\end{equation*}
which leads to
\begin{equation*}
\eps(n_1) n_1 < 0
\end{equation*}
Since we already know $\eps(n_1)$ this last inequality allows us to correctly choose the
sign of $n_1$. This same logic can be applied to all but the fifth face (which we will
deal with shortly). This leads to
\begin{spreadlines}{10pt}
\begin{align}
n_1^\mu &= -\sign{g^{44}}\frac{g^{4\mu}}{\sqrt{\abs{g^{44}}}}\label{eqn:StdNormalsA}\\
n_2^\mu &= -\sign{g^{33}}\frac{g^{3\mu}}{\sqrt{\abs{g^{33}}}}\label{eqn:StdNormalsB}\\
n_3^\mu &= -\sign{g^{22}}\frac{g^{2\mu}}{\sqrt{\abs{g^{22}}}}\label{eqn:StdNormalsC}\\
n_4^\mu &= -\sign{g^{11}}\frac{g^{1\mu}}{\sqrt{\abs{g^{11}}}}\label{eqn:StdNormalsD}
\end{align}
\end{spreadlines}
For the fifth face $(1234)$ the computations are slightly different. For this face the
vector $(04)$ is outward pointing, thus to ensure that $n_5^\mu$ is also outward
pointing we require
\begin{equation*}
\eps(n_5) n_{5\mu} (04)^\mu > 0
\end{equation*}
which leads to
\begin{equation*}
\eps(n_5) n_5 > 0
\end{equation*}
and thus
\begin{equation}
n_5^\mu = \frac{\sign{S}}{\sqrt{\abs{S}}}
          \left(g^{1\mu} + g^{2\mu} + g^{3\mu} + g^{4\mu}\right)
\label{eqn:StdNormalsE}
\end{equation}

Now we shall set about constructing two vectors $m_1^\mu$ and $m_2^\mu$ as partners to the
vectors $n_1^\mu$ and $n_2^\mu$. For $m_1^\mu$ we will require that it be a unit vector,
that it be orthogonal to both $(012)$ and $n_1^\mu$ and finally that it be oriented to
point away from $(012)$. Similar conditions will be imposed on $m_2^\mu$.

It is not hard to see that, apart from normalisation and orientation, the $m_i^\mu$ will
be of the form
\begin{spreadlines}{10pt}
\begin{align}
m_{1\mu} &= m_1\left(\delta_\mu^3 - \delta_\mu^4\frac{g^{34}}{g^{44}}\right)\\
m_{2\mu} &= m_2\left(\delta_\mu^4 - \delta_\mu^3\frac{g^{34}}{g^{33}}\right)
\label{eqn:StdTangentsA}
\end{align}
\end{spreadlines}
for some choice of numbers $m_1$ and $m_2$. From here on in the calculations are much like
those used for $n_i^\mu$. We first impose the normalisation conditions leading to
\begin{spreadlines}{12pt}
\begin{align}
\eps(m_1) &= \sign{g^{33} - \frac{g^{34}g^{34}}{g^{44}}}\\
\eps(m_2) &= \sign{g^{44} - \frac{g^{34}g^{34}}{g^{33}}}
\label{eqn:StdTangentsB}
\end{align}
\end{spreadlines}
while for the orientation we require
\begin{equation*}
\eps(m_1)m_{1\mu}(03)^\mu > 0\qquad\text{and }\qquad
\eps(m_2)m_{2\mu}(04)^\mu > 0
\end{equation*}
which leads to
\begin{spreadlines}{12pt}
\begin{align}
m_1 &= \eps(m_1)\left\vert{g^{33} - \frac{g^{34}g^{34}}{g^{44}}}\right\vert^{-1/2}\\
m_2 &= \eps(m_2)\left\vert{g^{44} - \frac{g^{34}g^{34}}{g^{33}}}\right\vert^{-1/2}
\label{eqn:StdTangentsC}
\end{align}
\end{spreadlines}

\subsection{Dot products}
\label{sec:DotProducts}

From the above definitions it is not hard to compute the following dot-products (valid only in
the standard frame).
\bgroup
\def\m{\phantom{-}}
\begin{align*}
m_1^\mu m_{1\mu} &= \m\sign{g^{33}-\frac{g^{34}g^{34}}{g^{44}}}\\[3pt]
m_2^\mu m_{2\mu} &= \m\sign{g^{44}-\frac{g^{34}g^{34}}{g^{33}}}\\[3pt]
n_1^\mu n_{1\mu} &= \m\sign{g^{44}}\\[8pt]
n_2^\mu n_{2\mu} &= \m\sign{g^{33}}\\[3pt]
m_1^\mu m_{2\mu} &=  -\sign{g^{33}g^{44}-g^{34}g^{34}}
                      \frac{g^{34}}{\vert g^{33}g^{44}\vert^{1/2}}\\[3pt]
n_1^\mu n_{2\mu} &= \m\sign{g^{33}g^{44}}
                      \frac{g^{34}}{\vert g^{33}g^{44}\vert^{1/2}}\\[3pt]
n_1^\mu m_{2\mu} &=  -\sign{g^{44}}
     \left\vert 1 - \frac{g^{34}g^{34}}{g^{33}g^{44}}\right\vert^{1/2}\\[3pt]
m_1^\mu n_{2\mu} &=  -\sign{g^{33}}
     \left\vert 1 - \frac{g^{34}g^{34}}{g^{33}g^{44}}\right\vert^{1/2}
\end{align*}
\egroup
These will be used when we develop explicit equations for the defect angles.

\section{Angles and boosts}
\label{sec:Angles}

One of the recurrent tasks in the Regge calculus is that of computing the angle between
pairs of vectors. This would be relatively straightforward if the metric was Euclidean.
However, for General Relativity, we require the metric to be Lorentzian and this
introduces some subtleties in the computations.

\subsection{The Euclidean plane}
\label{sec:EuclidAngle}

The following discussion may seem mundane and trivial but it does set the scene for the
slightly tricky aspects that we are about to encounter in the context of a Lorentzian plane.

Suppose that at a point in a Euclidean plane we have two unit-vectors. How do we measure the
angle between these vectors? The simple answer given in most elementary textbooks is that it
equals the length of the arc of the unit circle (centred on the given point) bounded by the
tips of the vectors. There are of course two arcs that join the tips and thus two possible
choices for the angle. A specific choice is usually made by other considerations (\eg we
could choose the arc that follows a counter clockwise direction, which in turn requires a
definition of clockwise but for the present audience we will take that as understood).

In the Lorentzian plane the set of unit-vectors at a point does not describe a closed path
but instead four branches of the unit-hyperbola. This posses a problem in extending the above
definition to the Lorentzian case, namely, that it is not clear how to measure the arc-length
between points that lie on different branches. Thus to avoid this problem we shall re-visit
the above definition of angles.

The set of all unit-vectors at a point in the Euclidean plane describes a unit-circle. Thus
we can establish a one-to-one mapping between the points on the unit-circle and the
unit-vectors at a point. A simple way to make this mapping explicit is to introduce
rectangular Cartesian coordinates. Let $\phi$ be a parameter that describes a typical point
on the unit circle and let $u(\phi)$ be the corresponding unit-vector with components
$(u^x(\phi),u^y(\phi))$. Then we have
\begin{equation}
u(\phi) = e_x u^x(\phi) + e_y u^y(\phi) = e_x\cos\phi + e_y\sin\phi
\label{eqn:EuclidAngle}
\end{equation}
where $0\leq\phi<2\pi$. Notice that as the unit-circle is traced out in a counter-clockwise
direction, $\phi$ increases from $0$ to $2\pi$. The mapping is one-to-one so a unique inverse
exists, which we will denote by $\phi(u)$.

Now let $\phi(u,v)$ be the counter clockwise angle from $u$ to $v$. Then we can define
$\phi(u,v)$ by
\begin{equation*}
\phi(u,v) := \phi(v) - \phi(u)
\end{equation*}
where $\phi$ on the right hand side is obtained by inverting equations
(\ref{eqn:EuclidAngle}). Note that this definition leads to the following properties for
angles, namely,
\begin{itemize}
\item $\phi(u,-u) = \pm\pi$.
\item $\phi(u,v) = - \phi(v,u)$.
\item $\phi(u,v) = \phi(u,w)+\phi(w,v)$ for any $u,v$ and $w$.
\item $\phi(u,v)$ is invariant with respect to rotations of $u$ and $v$.
\item If $u$ and $v$ are orthogonal then $\phi(u,v)=\pi/2$ or $3\pi/2$.
\item $\phi(u,v)$ is bounded.
\end{itemize}

The above definitions work well in a 2-dimensional space but for the Regge calculus we will be
working with 4-dimensions. How then do we generalise this definition to higher dimensions? The
simple answer is to use dot-products. Thus, given a flat Euclidean metric $g_{\mu\nu}$ and a
pair of unit-vectors $u$ and $v$, the angle $\phi(u,v)$ can be computed from
\begin{equation*}
\cos\phi(u,v) = g_{\mu\nu} u^\mu v^\nu
\end{equation*}
Is this definition equivalent to that given above, equation (\ref{eqn:EuclidAngle}), \ie will
the two equations yield the same angle $\phi(u,v)$ for any pair of unit vectors? Clearly the
answer is yes and the proof is trivial. Expand both vectors $u$ and $v$ on an orthonormal
basis. We are free to rotate the basis so that the vectors are of the form $u=e_x$ and
$v=e_x\cos\alpha+e_y\sin\alpha$. Upon substituting these into the right hand sides of both
definitions we see that their left hand sides agree, \ie they give the same angle, in this
case $\phi(u,v)=\alpha$.

\subsection{The Lorentzian plane}
\label{sec:LorentzBoost}

The preceding discussion is all rather simple and could easily be skipped over. However it
does serve to motivate how we propose to compute angles in a Lorentzian space.

Unlike the Euclidean case, the set of all possible unit-vectors at a point defines four
branches of a unit-hyperbola, which we have drawn in figure (\ref{fig:fig-01}) using the
parametric form
\begin{equation}
\def\M{\phantom{-}}
\def\m{-}
u(\phi) = e_t u^t(\phi) + e_x u^x(\phi) =
\begin{cases}
\M e_t\sinh\phi + e_x\cosh\phi&\hbox{in I}\\
\M e_t\cosh\phi + e_x\sinh\phi&\hbox{in II}\\
\m e_t\sinh\phi - e_x\cosh\phi&\hbox{in III}\\
\m e_t\cosh\phi - e_x\sinh\phi&\hbox{in IV}
\end{cases}
\label{eqn:LorentzBoost}
\end{equation}
where $u^x$ and $u^t$ are the components of a typical radial unit-vector,
$u(\phi)=e_t u^t(\phi) +e_x u^x(\phi)$.

Notice that in proceeding along the four branches in a counter-clockwise fashion the angle
$\phi$ does not increase monotonically. Indeed in regions II and IV the angle \emph{decreases}
from $+\infty$ to $-\infty$. Why then have we made this somewhat peculiar choice? Simply, it
is the only choice (apart from changing the sign of $\phi$ on all four branches) that ensures
that the angle between a pair of vectors is invariant with respect to boost transformations.
We will return to this point in just a moment.

We can use the above parametric form to suggest a mapping between points on the
unit-hyperbola and unit-vectors at a point. In this case though the mapping is not one-to-one
(to each $\phi$ there are four possible unit-vectors $u$), however, given a vector $u$ there
exists exactly one value of $\phi$ and thus it is meaningful to define $\phi$ as a function
of $u$. Thus as before we can define the angle between any pair of unit-vectors in the
Lorentzian plane by way of
\begin{equation*}
\phi(u,v) := \phi(v) - \phi(u)
\end{equation*}
where the right hand side is obtained by inverting equations (\ref{eqn:LorentzBoost}).

We now return to the point raised above concerning the non-monotonic behaviour of $\phi$. Had
we chosen a different orientation along the second branch, such as
\begin{equation*} 
  u(\phi) = e_t u^t(\phi) + e_x u^x(\phi) = e_t\cosh\phi - e_x\sinh\phi\quad\quad\hbox{in II}
\end{equation*} 
then the pair of vectors defined by $u=e_t\cosh1+e_x\sinh1$ in region II and
$v=e_t\sinh1+e_x\cosh1$ in region I would have $\phi(u)=-1$ and $\phi(v)=+1$. Thus $u$ and $v$
would be separated by an angle of $2$. But we recognise that $u$ and $v$ are the result of a
boost applied to the pair of vectors $e_t$ and $e_x$ respectively. And as the angle between
$e_t$ and $e_x$ is zero we see that the modification suggested above violates the boost
invariance of the angle between pairs of vectors.

If now we turn to dot-products then we find from the above parameterisations that
\begin{align*}
 \cosh\phi(u_{I},  v_{I}) &= g_{\mu\nu} u^\mu_{I}   v^\nu_{I}\\
 \sinh\phi(u_{II}, v_{I}) &= g_{\mu\nu} u^\mu_{II}  v^\nu_{I}\\
-\cosh\phi(u_{III},v_{I}) &= g_{\mu\nu} u^\mu_{III} v^\nu_{I}\\
-\sinh\phi(u_{IV}, v_{I}) &= g_{\mu\nu} u^\mu_{IV}  v^\nu_{I}
\end{align*}
where the subscripts on $u$ and $v$ indicate the quadrant in which the vector is defined.
Though this accounts for just 4 of the 16 possible combinations it is all that we need to
compute the angle between any pair of vectors. For example, for the vectors $u_{II}$ and
$w_{III}$ we can use $\phi(u_{II},w_{III}) = \phi(w_{III},v_{I}) - \phi(u_{II},v_{I})$. This
result follows directly from the previous definitions given for $\phi(u,v)$.

The following properties of $\phi(u,v)$ are readily obtained from the above definition
(compare these with the similar properties listed in section (\ref{sec:EuclidAngle}))
\begin{itemize}
\item $\phi(u,-u) = 0$.
\item $\phi(u,v) = - \phi(v,u)$.
\item $\phi(u,v) = \phi(u,w)+\phi(w,v)$ for any $u,v$ and $w$.
\item $\phi(u,v)$ is invariant with respect to boosts of $u$ and $v$.
\item If $u$ and $v$ are orthogonal (and non-null) then $\phi(u,v)=0$.
\item $\phi(u,v)$ is unbounded.
\end{itemize}
This last case can only arise when one or both of the vectors is null. We will deal with this
problem by simply excluding the case of null-vectors. Fortunately for most work in numerical
Regge calculus it is exceedingly unlikely that we would encounter a null-vector (when
computing defects). This is far from ideal but for the moment we will push on regardless
(while acknowledging the limitations of our description).

\section{Defect angles}
\label{sec:Defects}

The customary approach to computing defects is to use what one might loosely call a \emph{sum
of angles} method. This goes as follows. Suppose we have a 4-dimensional Euclidean lattice.
Pick any triangle $\sigma_2$ in the lattice. This triangle is surrounded by a closed loop of
4-simplices. Pick any one of these, call it $\sigma_4$. This 4-simplex will contain two
tetrahedral faces $\sigma_{3,1}$ and $\sigma_{3,2}$ that are themselves also attached to
$\sigma_2$. Now measure the angle subtended at $\sigma_2$ by this pair of tetrahedra, call it
$\phi_{12}$. Finally, repeat this for each of the remaining 4-simplices in the closed loop.
Then the defect on $\sigma_2$ is simply defined as $\theta= 2\pi - \sum \phi_{12}$. The
dihedral angles are computed (almost without exception) by way of
\begin{equation}
\sin\phi_{12} = \frac{4}{3} \frac{V_4 V_2}{V_{3,1}V_{3,2}}
\label{eqn:DihedralAngle}
\end{equation}
where $V_{3,1}$ and $V_{3,2}$ are the volumes of the two tetrahedral faces of $\sigma_4$ while
$V_4$ is the 4-volume of $\sigma_4$ and $V_2$ is the area of $\sigma_2$ (all of which can be
readily computed using the equations given in section (\ref{sec:Volumes})).

The above prescription is immediately applicable to a Euclidean simplicial lattice, however,
for a Lorentzian simplicial lattice some minor changes must be made. In particular, for a
spacelike $\sigma_2$ the right hand side of equation (\ref{eqn:DihedralAngle}) may have values
greater than one. This problem was dealt with by Wheeler \cite{wheeler:1964-01} and Sorkin
\cite{sorkin:1975-01} by allowing the dihedral angles to be complex valued. This in turn
required that the areas of the triangles also be complex valued in order that the Regge action
would remain real valued. Though this approach is correct it does require some care in
choosing the correct branch for the inverse sine function (again, see Sorkin
\cite{sorkin:1975-01} for full details).

We shall take a different (but equivalent) approach to Wheeler and Sorkin by working entirely
with real valued expressions. This is a minor advantage in any computer code as we would only
need to reserve half the memory required in the Wheeler and Sorkin approach (the defects, in
contrast to the dihedral angles, are always either purely real or purely imaginary and storing
such numbers as complex numbers is excessive).

In the following sections we will provide full details of two methods for computing the
defects, the first uses the sum of angles method while the second uses a parallel transport
method.

But before doing so we shall take the opportunity to make a small comment regarding a common
pictorial representation of defect angles in the Regge calculus. We will assume for this
discussion that we have a simple 2-dimensional lattice consisting of a set of triangles
attached to one vertex. A popular representation of such a lattice is to make a cut along one
of the radial edges so that triangles can be laid out flat in a 2-dimensional space. A typical
example, for a Euclidean lattice, is shown in figure (\ref{fig:fig-03}). The shaded wedge is
the part of the space not covered by the lattice. Also, to recover the lattice, we must
identify the radial edges of the wedge. In this picture we have also included the unit circle
so that we can easily measure the angle subtended by the wedge and clearly this angle is
exactly equal to the defect angle of the lattice. It should also be clear that the cut can be
made along any radial line starting from the bone. Choosing some other radial leg produces a
similar diagram but with the wedge rotated about the bone. The flexibility to locate the cut
anywhere in the lattice does not apply for a Lorentzian lattice. This is easily seen by
reference to figures (\ref{fig:fig-04}--\ref{fig:fig-06}). In figure (\ref{fig:fig-04}) we see
that the defect angle would be positive yet in figure (\ref{fig:fig-05}) we have a negative
defect (the angle decreases along this branch when traversed in the counter-clockwise
direction). Finally, in figure (\ref{fig:fig-06}) we show an impossible construction -- the
edges of the wedge can not be identified because their tangent vectors have differing
signatures.

\subsection{Sum of angles}
\label{sec:SumOfAngles}

The definition here is very simple. The defect $\theta_{\sigma_2}$ on a typical bone
$\sigma_2$ is defined by
\begin{equation}
\theta_{\sigma_2} =
\begin{cases}
\dsp\quad 2\pi - \sum_{\sigma_4(\sigma_2)}\>
   \left(\phi_{12}\right)_{\sigma_4}&\text{for $\sigma_2$ timelike}\\
\noalign{\vskip10pt}
\dsp\quad \phantom{2\pi} - \sum_{\sigma_4(\sigma_2)}\>
   \left(\phi_{12}\right)_{\sigma_4}&\text{for $\sigma_2$ spacelike}
\end{cases}
\label{eqn:DefectSumOfAngles}
\end{equation}
where the sum includes all of the 4-simplices $\sigma_4$ attached to the bone $\sigma_2$. The
dihedral angles $\phi_{12}$ can be computed using equation
(\ref{eqn:DihedralAngle}) taking due care for the complex numbers that might arise. On the
other hand we can work solely with real numbers by combining our definitions of angles given
in section (\ref{sec:Angles}) together with the dot-products of the normal and tangent vectors
(as defined in section (\ref{sec:Normals})). Then for a \emph{timelike} bone we find
\begin{equation}
   \phi_{12} = \arccos\left(m_1^\mu m_{2\mu}\right)
   \label{eqn:AngleDotProductA}
\end{equation}
while for a \emph{spacelike} bone we find
\begin{equation}
   \phi_{12} = \sign{m_1^\mu m_{1\mu}}\arcsinh\rho_{12}
   \label{eqn:AngleDotProductB}
\end{equation}
where $\rho_{12}$ is computed from
\begin{equation}
   \rho_{12} = 
   \begin{cases}
       \sign{n_1^\nu m_{2\nu}}m_1^\mu m_{2\mu}&\text{when\ }\abs{m_1^\mu m_{2\mu}}<\abs{n_1^\mu m_{2\mu}}\\[5pt]
       \sign{m_1^\nu m_{2\nu}}n_1^\mu m_{2\mu}&\text{in all other cases}
   \end{cases}
\end{equation}
Recall that the dot-products in the standard frame are given in section
(\ref{sec:DotProducts}).

\subsection{Parallel transport}
\label{sec:ParallelTrans}

Here we will use a commonly quoted but rarely used method for computing a defect angle. After
parallel transporting a vector around a simple closed loop the defect angle is computed as the
rotation angle (or boost) between the initial and final vectors. If we denote the initial and
final vectors by $v$ and $v'$ respectively, then we expect a transformation of the form
\begin{equation*}
v' = Rv
\end{equation*}
where $R$ is a $4\times4$ rotation (or boost) matrix (this matrix will act trivially on
vectors parallel to the bone).

We will construct $R$ by taking account of the separate transformations that arise, first from
the parallel transport within a 4-simplex, and second from the passage from one 4-simplex to
the next. In each case we will employ suitably chosen basis vectors local to each 4-simplex.
The successive stages of the transformation will be recorded as a matrix of scalars with
respect to the local basis.

We begin by considering the transformation associated with the parallel transport within
a typical 4-simplex, $(ijk12)$. The section of the closed loop that passes
through this 4-simplex will have entry and exit points on distinct\footnote{We exclude the
case where the entry and exit faces are one and the same, \eg when the path doubles back on
itself.} faces of that 4-simplex. Suppose that the entry face is $(ijk1)$ and the exit face is
$(ijk2)$. At the entry point we will use the (non-orthogonal) tetrad
\begin{align*}
e_1&=(ij)& 
e_2&=(ik)& 
e_3&=n_1^\mu e_\mu& 
e_4&=m_1^\mu e_\mu
\intertext{while at the exit we will use}
e_1'&=(ij)& 
e_2'&=(ik)& 
e_3'&=n_2^\mu e_\mu& 
e_4'&=m_2^\mu e_\mu
\end{align*}
where the $e_\mu$ are the coordinate basis vectors in the standard frame and $n_1^\mu$,
$n_2^\mu$, $m_1^\mu$ and $m_2^\mu$ are defined in section (\ref{sec:Normals}). Then the
parallel transport of the vector $v$ along the path within this 4-simplex is described by
\begin{equation*}
v = v^a e_a = v^{'a} e_a'
\end{equation*}
where the $v^a$ and $v^{'a}$ are the components of $v$ onto the
respective tetrads. We also know that the entry and exit tetrads are related
by a simple linear transformation of the form
\begin{align}
e_1 &= e_1' &
e_2 &= e_2' &
e_3 &= T_3^3 e_3' + T_3^4 e_4' &
e_4 &= T_4^3 e_3' + T_4^4 e_4'
\label{eqn:BasisPT}
\end{align}
for some set of numbers $T_a^b$ (which we will compute explicitly in the following section).
This in turn implies a similar linear transformation on the components, $v^{'a}$ and
$v^a$, 
\begin{align*}
v^{'1} &= v^1 &
v^{'2} &= v^2 &
v^{'3} &= T_3^3 v^3 + T_4^3 v^4 &
v^{'4} &= T_3^4 v^3 + T_4^4 v^4
\end{align*}
This completes the transport of the vector within one 4-simplex. Now we must take account of
the changes that occur as the vector is transported across the junction between a pair of
4-simplices. This is very easy to do, we need only recognise that the normal vector of the
exit face (of the current 4-simplex) is oppositely oriented to the normal vector on the entry
face (of the next 4-simplex). This amounts to the following transformation
\begin{equation*}
v^a \mapsto S^a_b v^b
\end{equation*}
where $S = \diag(1,1,-1,1)$.

We are now ready to compute the transformation matrix $R$ for the closed path around the bone.
Let $T_m$ be the matrix built from the $T_a^b$ associated with the $m^{th}$ 4-simplex, then we
have
\begin{equation}
R =\left(ST_n\right)\left(ST_{n-1}\right)\left(ST_{n-2}\right)\cdots\left(ST_1\right)
\label{eqn:MatrixPT}
\end{equation}
What can we say about the form of $R$? We know that after one loop of the bone any vector
initially parallel to the bone will be returned unchanged while any vector orthogonal to the
bone will be subjected to a rotation (for a timelike bone) or a boost (for a spacelike bone).
Thus we expect $R$ to be of the form
\begin{align}
R &= 
\begin{bmatrix}
\ 1&0&0&0\\
0&1&0&0\\
0&0&\cos\beta&-\sin\beta\\
0&0&\sin\beta&\phantom{-}\cos\beta
\end{bmatrix}\qquad\text{for a timelike bone}
\label{eqn:StandardMatrixPTa}
\intertext{and}
R &= 
\begin{bmatrix}
\ 1&0&0&0\\
0&1&0&0\\
0&0&\cosh\beta&\sinh\beta\\
0&0&\sinh\beta&\cosh\beta
\end{bmatrix}\qquad\text{for a spacelike bone}
\label{eqn:StandardMatrixPTb}
\end{align}
for some number $\beta$. The question now is, what is the relationship between $\beta$ and the
defect angle $\theta$? We will answer this question by requiring that the defect angle
calculated by the parallel transport method agrees with that given by the standard sum of
angles method.

We begin with the simple case of a timelike bone with a defect angle $\theta$. To bring the
sum of angles method into the picture we will use the representations introduced in section
(\ref{sec:Defects}). Thus in figure (\ref{fig:fig-08}) we have a timelike bone with a defect
angle $\theta$. We have also included in that figure the results of parallel transporting a
pair of vectors $n,m$ along a closed path around the bone. Note that the shaded region is
excluded from the figure and that the radial lines should be identified. We have also
suppressed the two dimensions parallel to the bone (there is no useful information contained
in those dimensions). Equally, the figure could be viewed as an isometric mapping of the
2-dimensional subspace orthogonal to the bone into a subset of $R^2$ with a Euclidean metric.
After one loop of the bone we obtain the pair of vectors $n',m'$ and these are to be compared
with the original vectors $n'',m''$. Evidently this entails a rotation through an angle
$\beta$ and so we deduce that the defect is given by $\theta=\beta$.

A little more care is required for the case of a spacelike bone. As we shall soon see, the
entries in the matrix $R$ may depend upon which 4-simplex is chosen to start the loop around
the bone. Consider the example shown in figure (\ref{fig:fig-09}). This is very similar to the
previous figure with two important differences, first we are now working with a Lorentzian
metric (in this 2-space) and second we must now speak of boost parameters (as opposed to
rotation angles). Thus the skewed appearance of $n,m$ in the figure is simply a consequence of
using a Lorentzian metric. By inspection we can see that the boost parameter $\beta$ (for the
boost from $n'',m''$ to $n',m'$) is negative while the defect is positive. Thus in this
instance we have $\theta=-\beta$. Now turn to figure (\ref{fig:fig-10}). This too is also for
the case of a spacelike bone but here we have made the cut at a position which has $n$ as a
spacelike vector (in the previous case $n$ was a timelike vector). Again we see that the boost
parameter is negative but on this occasion we have a negative defect, thus we have $\theta =
+\beta$.

This result can also be understood by purely algebraic means. Consider the transformation
matrices that would arise by completing the parallel transport around a loop by starting in
distinct 4-simplices. We expect that the defects computed from these matrices must all agree
(they describe the parallel transport around the same bone). In figure (\ref{fig:fig-07}) we
have drawn the first few 4-simplices attached to the bone $(ijk)$. On each 4-simplex we have
drawn their particular entry basis vectors (and for simplicity we have not drawn them as
skewed vectors). Let $R_1$ be the parallel transport matrix associated with $(ijk12)$, with
$R_2$ for $(ijk23)$ and so on. Suppose that $n_1$ and $n_2$ are timelike while $n_5$ is
spacelike. Let $P_{12}$ be the transformation matrix that maps $n_1,m_1$ into $n_2,m_2$. Then
it follows that
\begin{equation*}
R_1 = P^{-1}_{12} R_2 P_{12}
\end{equation*}
However, since $n_1$ and $n_2$ are both timelike we see that $P_{12}$ must be a pure boost.
Furthermore, $P_{12}$ and $R_2$ act in the same 2-space and thus they commute, so we have
\begin{equation*}
R_1 = P^{-1}_{12} R_2 P_{12} = P^{-1}_{12}P_{12} R_2 = R_2
\end{equation*}
which leads to $\beta_1 = \beta_2$.
If we attempt the same calculation for the simplices $(ijk12)$ and $(ijk56)$ we must take
account of the fact that $n_1$ is timelike while $n_5$ is spacelike. The map $P_{15}$ from
$n_1,m_1$ to $n_5,m_5$ will now be of the form
\begin{equation*}
P_{15} = B_{15} Q
\end{equation*}
where $B_{15}$ is a pure boost and (if we ignore the two dimensions parallel to the bone)
\begin{equation*}
Q = 
\begin{bmatrix}
0&1\\-1&0
\end{bmatrix}
\end{equation*}
This leads to
\begin{align*}
R_1 &= P^{-1}_{15} R_5 P_{15}\\
    &= Q^{-1}B^{-1}_{15} R_5 B_{15}Q\\
    &= Q^{-1} R_5 Q
\end{align*}
and thus $\beta_1 = - \beta_5$.

So we now have a simple answer to the earlier question, how is $\theta$ related to $\beta$,
\begin{equation}
\theta = 
\begin{cases}
\phantom{\eps(n_1)}-\arcsin\left(R_{3}{}^{4}\right)&\text{for a timelike bone}\\
\phantom{-}\eps(n_1)\arcsinh\left(R_{3}{}^{4}\right)&\text{for a spacelike bone}
\end{cases}
\label{eqn:FinalDefectPT}
\end{equation}

\subsubsection{The standard frame}

We can compute the $T_a^b$ entries in each transformation matrix $T_m$ by taking
appropriate scalar products of (\ref{eqn:BasisPT}). This leads to
\begin{align*}
T_3^3 &=\frac{n_{1\mu}n_2^\mu}{n_{2\nu}n_2^\nu}&    
T_3^4 &=\frac{n_{1\mu}m_2^\mu}{m_{2\nu}m_2^\nu}\\   
T_4^3 &=\frac{m_{1\mu}n_2^\mu}{n_{2\nu}n_2^\nu}&    
T_4^4 &=\frac{m_{1\mu}m_2^\mu}{m_{2\nu}m_2^\nu}     
\end{align*}
with all other $T_a^b=0$ except $T_1^1=T_2^2=1$.
In the standard frame we find that
\bgroup
\def\m{\phantom{-}}
\begin{spreadlines}{10pt}
\begin{align}
T_3^3 &=\m\sign{g^{44}}g^{34}\left\vert g^{33}g^{44}\right\vert^{-1/2}
\label{eqn:SimpleTija}\\
T_4^4 &=-\sign{g^{33}}g^{34}\left\vert g^{33}g^{44}\right\vert^{-1/2}
\label{eqn:SimpleTijb}\\
T_4^3 &=-\left\vert 1-\frac{g^{34}g^{34}}{g^{33}g^{44}} \right\vert^{1/2}
\label{eqn:SimpleTijc}\\
T_3^4 &=-\sign{1-\frac{g^{34}g^{34}}{g^{33}g^{44}}}
        \left\vert 1-\frac{g^{34}g^{34}}{g^{33}g^{44}} \right\vert^{1/2}
\label{eqn:SimpleTijd}
\end{align}
\end{spreadlines}
\egroup

From the above it is not hard to show that $\det\left(ST_m\right) = 1$.

In summary, we would use equations (\ref{eqn:SimpleTija}--\ref{eqn:SimpleTijd}) to compute the
successive $T_m$, these would then be used to compute $R$ from (\ref{eqn:MatrixPT}) and
finally equation (\ref{eqn:FinalDefectPT}) would be used to compute the defect.

There is one further subtlety in the transformation process that we neglected to mention. For
a general path (not necessarily the closed path around a bone) the 2-simplex that lies at the
intersection of the entry and exit faces will change along the path. This has not been
accounted for in the above analysis but the adaptions required are simple and follow along
lines similar to that given above. Consider two 4-simplices $(ijk12)$ and $(ijk23)$. Suppose
the entry and exit faces for the first 4-simplex are $(ijk1)$ and $(ijk2)$ while in the second
4-simplex the entry and exit faces are $(ijk2)$ and $(jk23)$. Thus upon entry to the second
4-simplex the tetrad will be
\begin{align*}
e_1&=(ij) &
e_2&=(ik) &
e_3&=n_1^\mu e_\mu &
e_4&=m_1^\mu e_\mu
\intertext{while at the exit we might use}
e_1'&=(jk) &
e_2'&=(j2) &
e_3'&=n_4^\mu e_\mu &
e_4'&=m_4^\mu e_\mu
\end{align*}
and this will impose a further transformation on the tetrad components prior to leaving the
second simplex. We choose not to give the full details here as they are not required for the
case of most interest, namely, the computation of the defect angle.

\section{Derivatives of defects}
\label{sec:Derivs}

One particularly attractive feature of the standard frame is that it allows us to compute
the derivatives of the defects at virtually no extra computational cost beyond that
required for the defects. In short, we get the derivatives for free.

We begin by considering one 4-simplex $\sigma_4:=(ijk12)$ and focusing on the angle
$\phi_{12}$ between the pair of 3-simplices $\sigma_3(1):=(ijk1)$ and $\sigma_3(2):=(ijk2)$.
Suppose for the moment that $\sigma_2:=(ijk)$ is timelike then with the orientations for
$n^\mu$ and $m^\mu$ chosen as per figure (\ref{fig:fig-02}) we have
\begin{align*}
m_{2\mu} &= - n_{1\mu}\sin\phi_{12} + m_{1\mu}\cos\phi_{12}\\[5pt]
n_{2\mu} &= - n_{1\mu}\cos\phi_{12} - m_{1\mu}\sin\phi_{12}
\end{align*}
If some small changes are now made to the lattice (\eg by small changes in the leg-lengths)
then we must have
\bgroup
\def\m{\phantom{-}}
\begin{align*}
\delta m_{2\mu} &= \m n_{2\mu}\delta\phi_{12}
                 - \delta n_{1\mu}\sin\phi_{12} + \delta m_{1\mu}\cos\phi_{12}\\[5pt]
\delta n_{2\mu} &= - m_{2\mu}\delta\phi_{12} 
                 - \delta n_{1\mu}\cos\phi_{12} - \delta m_{1\mu}\sin\phi_{12}
\end{align*}
\egroup
and thus
\begin{equation*}
  2\delta\phi_{12} =  n_2^\mu\delta m_{2\mu} - m_2^\mu\delta n_{2\mu}
                    + m_1^\mu\delta n_{1\mu} + n_1^\mu\delta m_{1\mu}\\[5pt]
\end{equation*}
This result applies in all frames but when restricted to the standard
frame it takes on a particularly simple form. In the standard frame we have
\begin{alignat*}{3}
m_{1\mu} &= m_1\left(\delta^3_\mu+\alpha_1\delta^4_\mu\right)&&\qquad\qquad&n_{1\mu} 
  &= n_1\delta^4_{\mu}\\
m_{2\mu} &= m_2\left(\delta^4_\mu+\alpha_2\delta^3_\mu\right)&&&n_{2\mu} 
  &= n_2\delta^3_{\mu}
\end{alignat*}
where the $m_i$ and $n_i$ are normalisation factors while $\alpha_i$ is chosen so that
$0=n_{i\mu}m_i^\mu$. Thus we have
\begin{equation*}
0=m_i^\mu\delta n_{i\nu}\qquad\text{for }i=1,2
\end{equation*}
while from $0=n_{i\mu}m_i^\mu$, $0=n_i^\mu m_{i\mu}$ and
$0=g_{\mu\nu}n^\mu_i m^\mu_i$ we also have
\begin{equation*}
n_i^\mu \delta m_{i\mu} = \delta g_{\mu\nu} n^\mu_i m^\mu_i\qquad\text{for }i=1,2
\end{equation*}
which leads to the following simple equation for the variations in the angle
\begin{equation*}
2\delta\phi_{12} = \left(m^\mu_2 n^\nu_2 + m^\mu_1 n^\nu_1\right) \delta g_{\mu\nu}
\end{equation*}
An almost identical analysis can be applied in the case of a spacelike bone. It begins with
\begin{align*}
m^\mu_2 &= - n^\mu_1\sinh\phi_{12} + m^\mu_1\cosh\phi_{12}\\[5pt]
n^\mu_2 &= - n^\mu_1\cosh\phi_{12} + m^\mu_1\sinh\phi_{12}
\end{align*}
and leads to
\begin{equation*}
2\delta\phi_{12} = - \left(m^\mu_2 n^\nu_2 + m^\mu_1 n^\nu_1\right) \delta g_{\mu\nu}
\end{equation*}

To compute the derivative of the defect $\theta$ on $\sigma_2$ we need only combine the
above results with the equation (\ref{eqn:DefectSumOfAngles}) for the defect to obtain
\begin{equation}
\left( \pderiv(\theta,L^2) \right)_{\sigma_2}
=  \frac{1}{2}\eps(\sigma_2)\sum_{\sigma_4(\sigma_2)} 
   \left(\left(m^\mu_2 n^\nu_2 + m^\mu_1 n^\nu_1\right) 
   \pderiv(g_{\mu\nu},L^2)\right)_{\sigma_4}
\label{eqn:DefectDerivs}
\end{equation}
where the sum includes each of the 4-simplices attached to $\sigma_2$, $L^2$ is a typical
(squared) leg-length and $\eps(\sigma_2) = -1$ for a timelike bone and $+1$ for a spacelike
bone.

The utility of this equation should not be overlooked. The partial derivatives of the metric
in the standard frame are simple numbers such as $1,\pm1/2$ and zero and thus there is no
extra cost (of any importance) in computing the Jacobian of the Regge equations in situ with
the equations themselves. All of the terms in this equation are already in use during the
computation of the defects. This is a significant advantage when it comes to solving the
Regge equations by standard numerical methods, having the Jacobian at hand at no cost is a
great bonus. Of course the Jacobian could be computed by finite differences but that would be
extremely expensive. Each leg would need to be varied in turn and the corresponding changes
in the defects recorded. If we suppose that, on average, each defect depends on $N$ legs,
then this finite-difference process will increase the computational cost of the defects by a
factor of $N$, a significant expense (a bare minimum for $N$ is 10, the 10 legs of one
4-simplex).

\section{Timing}
\label{sec:Timing}

The stated aim of this paper is to present efficient algorithms for computing the defects and
their derivatives. Here we will present the results of some simple tests. We chose two simple
lattices each consisting of one bone, a timelike bone in one lattice and a spacelike bone in
the other. The leg lengths were assigned by choosing a well known metric (\eg Schwarzschild,
plane waves, Kasner etc.) and using a geodesic integrator to compute the lengths of short
geodesics (see \cite{brewin:2000-01} for full details). The results presented here are a
measure of the arithmetic complexity of each algorithm. Thus it is reasonable to expect that
the our results are largely independent of the details of the lattice (\eg the size, number
and location of each 4-simplex in the parent spacetime).

In our first test we computed the defects using the sum of the angles method as well as the
parallel transport method. The angles required in the sum of angles method can be computed in
two ways, by the computing ratios of volumes or by dot products of normal vectors. In table
(\ref{tbl:ResultsDefects}) we have listed the cpu times to compute the defects for each of
these methods. The times are expressed as ratios relative to the standard algorithm (\ie where
angles are computed by ratios of volumes).
\bgroup
\begin{table}[ht]
\def\H{\vrule height 14pt depth  7pt width 0pt} 
\def\m{\vrule height  0pt depth 10pt width 0pt} 
\def\M{\vrule height 15pt depth 10pt width 0pt}
\begin{center}
\begin{tabular}{ccccccc}
\hline
\H&Algorithm&&Equations&&CPU time&\\
\hline
\M&Ratios of volumes&&(\ref{eqn:DihedralAngle})&&\TimeDefectTraditional\\
\m&Dot products&&(\ref{eqn:AngleDotProductA},\ref{eqn:AngleDotProductB})&&\TimeDefectParallel\\
\m&Parallel transport&&(\ref{eqn:FinalDefectPT})&&\TimeDefectStandard\\
\hline
\end{tabular}
\end{center}
\caption{%
CPU times for three algorithms used to compute the defects.}
\label{tbl:ResultsDefects}
\end{table}
\egroup
This shows that the dot product and parallel transport algorithms are comparable in speed and
that both are nearly twice as fast as the standard algorithm. This is a good start.

Now we turn to the computation of the derivatives of the defects. Here we will work solely
with the dot product algorithm (the angles by ratios of volumes algorithm is too slow while
the parallel transport method runs neck and neck with the dot product method).

In this test we compute the cpu time to compute the derivatives of the defects with respect to
the three legs of the bone. We do so using two algorithms, the first uses the exact expression
(\ref{eqn:DefectDerivs}) derived in section (\ref{sec:Derivs}) and the second uses centred
finite differences on equations (\ref{eqn:AngleDotProductA},\ref{eqn:AngleDotProductB}). The
results, normalised against the cpu time to compute the defect, are listed in table
(\ref{tbl:ResultsDerivs}).
\bgroup
\begin{table}[ht]
\def\H{\vrule height 14pt depth  7pt width 0pt} 
\def\m{\vrule height  0pt depth 10pt width 0pt} 
\def\M{\vrule height 15pt depth 10pt width 0pt}
\begin{center}
\begin{tabular}{ccccccc}
\hline
\H&Algorithm&&Equations&&CPU time&\\
\hline
\M&Exact derivatives&&(\ref{eqn:DefectDerivs})&&\TimeDerivsExact\\
\m&Finite differences&&(\ref{eqn:AngleDotProductA},\ref{eqn:AngleDotProductB})&&\TimeDerivsApprx\\
\hline
\end{tabular}
\end{center}
\caption{%
CPU times to compute the three derivatives of the defect.}
\label{tbl:ResultsDerivs}
\end{table}
\egroup
This clearly shows that there is very little overhead when computing the derivatives using the
exact equations (\ref{eqn:DefectDerivs}). In contrast the finite difference method is close to
six times slower. This figure is easily understood -- for each of the three legs we have to
compute the defect twice, hence a total of six defect computations. Now consider a typical
bone in a typical simplicial lattice. The defect on this bone will be depend on large number
of legs, at least 10 and possibly upwards of 50. The above calculations suggest that computing
all the derivatives in this case could be as bad as 100 times slower than by the exact
equations. This would be so slow as to be impractical (assuming no other part of the code
dominated the computations).

\section{Cross checking}
\label{sec:CrossCheck}

It is a common (and regrettable) fact that errors of many kinds (numerical, coding, logical)
do creep into computer programs. The wise programmer will employ as many tests as they can to
validate their code. For us we have two identities that can be usefully employed, Stokes'
theorem for one 4-simplex,
\begin{equation}
0 = \sum_{i=1}^5 \eps(n_i) n_{i\mu} V_{3i}
\label{eqn:StokesThm}
\end{equation}
which can be used to test that the unit normals are correctly oriented and Regge's identity
\begin{equation}
0 = \sum_{\sigma_2} \left(A \pderiv(\theta,L^2)\right)_{\sigma_2}
\label{eqn:ReggeIdentity}
\end{equation}
which can be used to check the correctness of the defect angles (by employing 2nd order
finite differences and looking for 2nd order convergence of the right hand side to zero).


\subsection{Stokes' theorem}

A typical 4-simplex, say $(01234)$, has five faces. In section (\ref{sec:Normals}) we proposed
a simple form for the unit normal to each face from which it follows that
\begin{equation*}
0 = - \frac{n_{5\mu}}{n_5} + \sum_{i=1}^{4} \frac{n_{i\mu}}{n_i}
\end{equation*}
Out of this equation we can now construct a Stokes' theorem for one 4-simplex. Consider the
first face, say $(0123)$, and its outward normal vector in the form $n_{1\mu} =
n_1\delta^4_{\mu}$. Upon substituting this into equation (\ref{eqn:FourVol}) we find
\begin{equation*}
4V_4 = \vert n_1\vert V_{31} = \eps(n_1) n_1 V_{31}
\end{equation*}
where $V_4$ is the 4-volume of the 4-simplex and $V_{31}$ is the 3-volume of the first face.
Clearly we can repeat this for each of the remaining faces leading to
\begin{equation*}
4V_4 =   \eps(n_1) n_1 V_{31}
     =   \eps(n_2) n_2 V_{32}
     =   \eps(n_3) n_3 V_{33}
     =   \eps(n_4) n_4 V_{34}
     = - \eps(n_5) n_5 V_{35}
\end{equation*}
This can be used to eliminate the $n_i$ in the first equation of this section, the result is
\begin{equation*}
0 = \sum_{i=1}^5 \eps(n_i) n_{i\mu} V_{3i}
\end{equation*}

\subsection{Regge's identity}

Here we will use equation (\ref{eqn:DefectDerivs}) for the derivatives of the defect to obtain
a simple proof of Regge's identity (\ref{eqn:ReggeIdentity}). We being by forming a weighted
sum of equation (\ref{eqn:DefectDerivs}) over all of the bones in the interior of the lattice
\begin{equation*}
 2\sum_{\sigma_2} \left( A \pderiv(\theta,L^2) \right)_{\sigma_2}
= \sum_{\sigma_2} \sum_{\sigma_4(\sigma_2)} 
  \left(\eps A\right)_{\sigma_2}
  \left(\left(m^\mu_2 n^\nu_2 + m^\mu_1 n^\nu_1\right) 
  \pderiv(g_{\mu\nu},L^2)\right)_{\sigma_4}
\end{equation*}
This sum contains contributions from each 4-simplex and its component sub-simplices (in
particular the faces and the bones). Upon careful inspection we see that we can re-order the
sums as follows
\begin{equation*}
 2\sum_{\sigma_2} \left( A \pderiv(\theta,L^2) \right)_{\sigma_2}
= \sum_{\sigma_4} \sum_{\sigma_3(\sigma_4)} \sum_{\sigma_2(\sigma_3)} 
     \left(\eps A m^\mu\right)_{\sigma_2} 
     \left( n^\nu \right)_{\sigma_3}
     \left(\pderiv(g_{\mu\nu},L^2)\right)_{\sigma_4}
\end{equation*}
Finally, by applying Stokes' theorem to each $\sigma_3$ we see that the innermost sum vanishes
and thus we recover Regge's identity (\ref{eqn:ReggeIdentity}).

\section{Historical note}
\label{sec:History}
The seeds of the theory that would later be known as the Regge calculus were sown in a
barber's shop in Princeton sometime in 1959 \cite{hartle:1985-01}. Tullio Regge, then a PhD
student working with John Wheeler, sat in the barber's chair and stared into the mirrors that
covered the four walls and ceiling of the shop. In these mirrors he saw the barber shop
repeated over and over and this led him to muse, while the barber went about his business,
that perhaps the universe might be built from, or approximated by, a large array of
interconnected cells. In 1961 Regge published his seminal paper, \emph{General Relativity
without coordinates}. The subject was championed by John Wheeler in \emph{Groups Relativity
and Topology}, who coined the title \emph{Regge calculus}.

\clearpage

\captionsetup{margin=0pt,font=small,labelfont=bf}

\def\Figure#1#2{%
\centerline{%
\includegraphics[width=#1\textwidth]{#2}}
\vskip0.5cm}

\def\FigPair#1#2{%
\centerline{%
\includegraphics[width=0.6\textwidth]{#1}\hfill%
\includegraphics[width=0.6\textwidth]{#2}}
\vskip0.5cm}

\def\FigQuad#1#2#3#4{%
\centerline{%
\includegraphics[width=0.6\textwidth]{#1}\hfill%
\includegraphics[width=0.6\textwidth]{#2}}%
\centerline{%
\includegraphics[width=0.6\textwidth]{#3}\hfill%
\includegraphics[width=0.6\textwidth]{#4}}
\vskip0.5cm}


\begin{figure}[t]
\Figure{1.0}{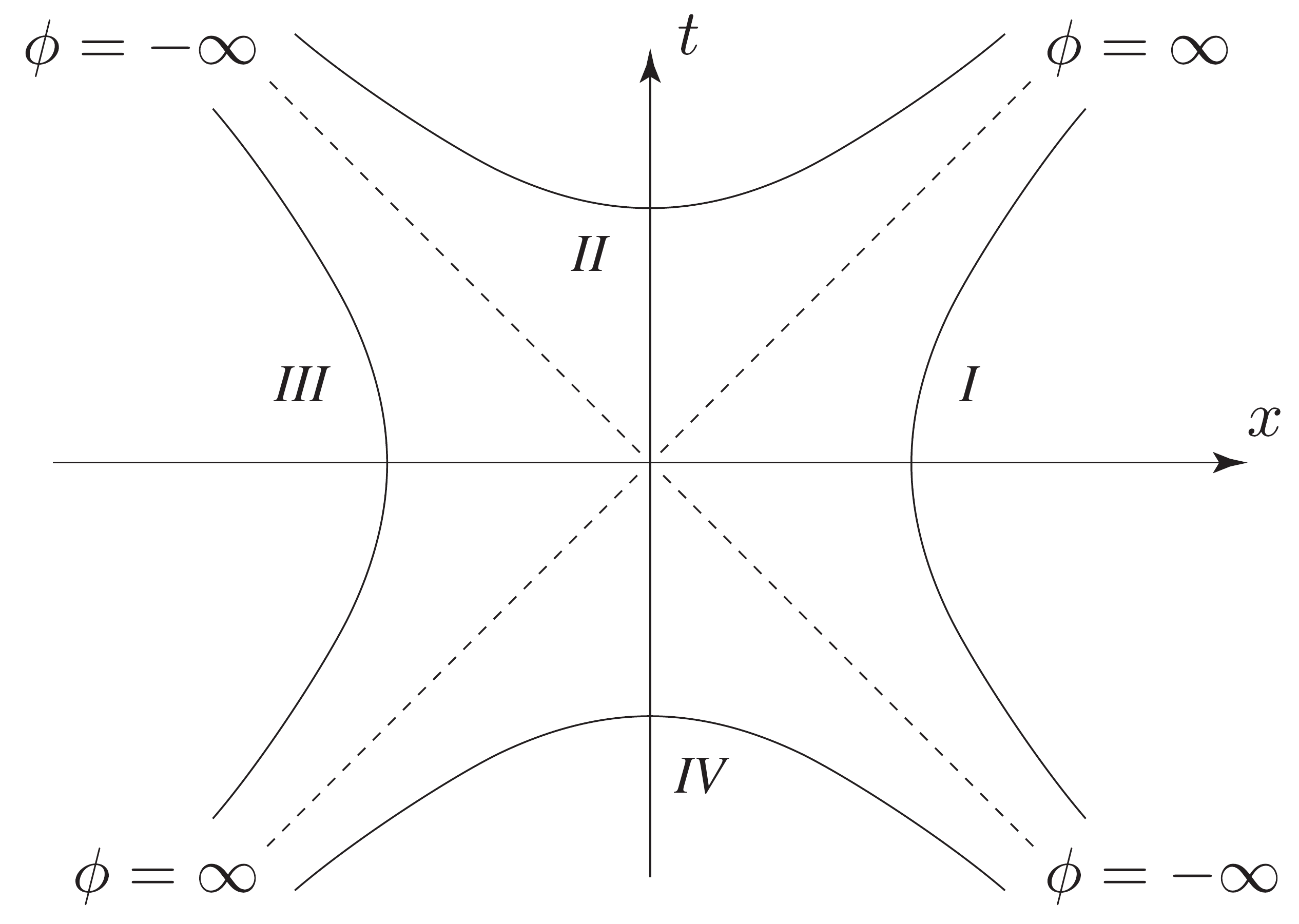}
\caption{\normalfont%
This figure shows our choice of angle in a 2-dimensional Lorentzian space. Note that the angle
does not increase monotonically as we traverse the branches in a counter-clockwise direction.}
\label{fig:fig-01}
\end{figure}

\begin{figure}[t]
\Figure{0.75}{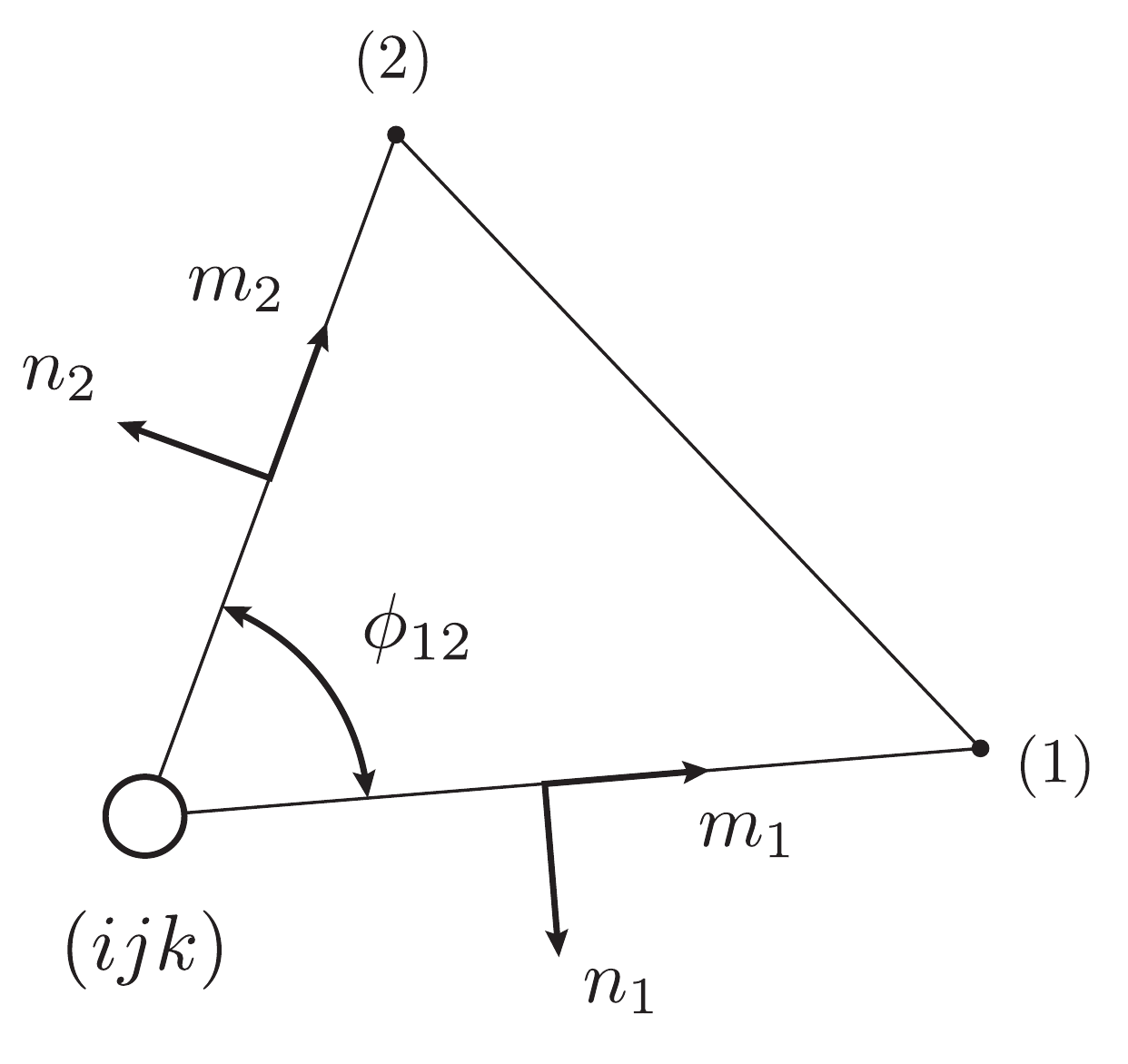}
\caption{\normalfont%
A typical 4-simplex $(ijk12)$. Here we show the outward pointing unit-vectors $n_1$, $n_2$ and
the corresponding unit tangent vectors $m_1$ and $m_2$. These vectors are used to compute the
dihedral angle $\phi_{12}$. They are also used as basis vectors for the parallel transport of
vectors around the bone $(ijk)$. The small circle is our way of noting that $(ijk)$ is a
2-simplex (whereas vertices are drawn as a solid dot).}
\label{fig:fig-02}
\end{figure}

\begin{figure}[t]
\Figure{0.75}{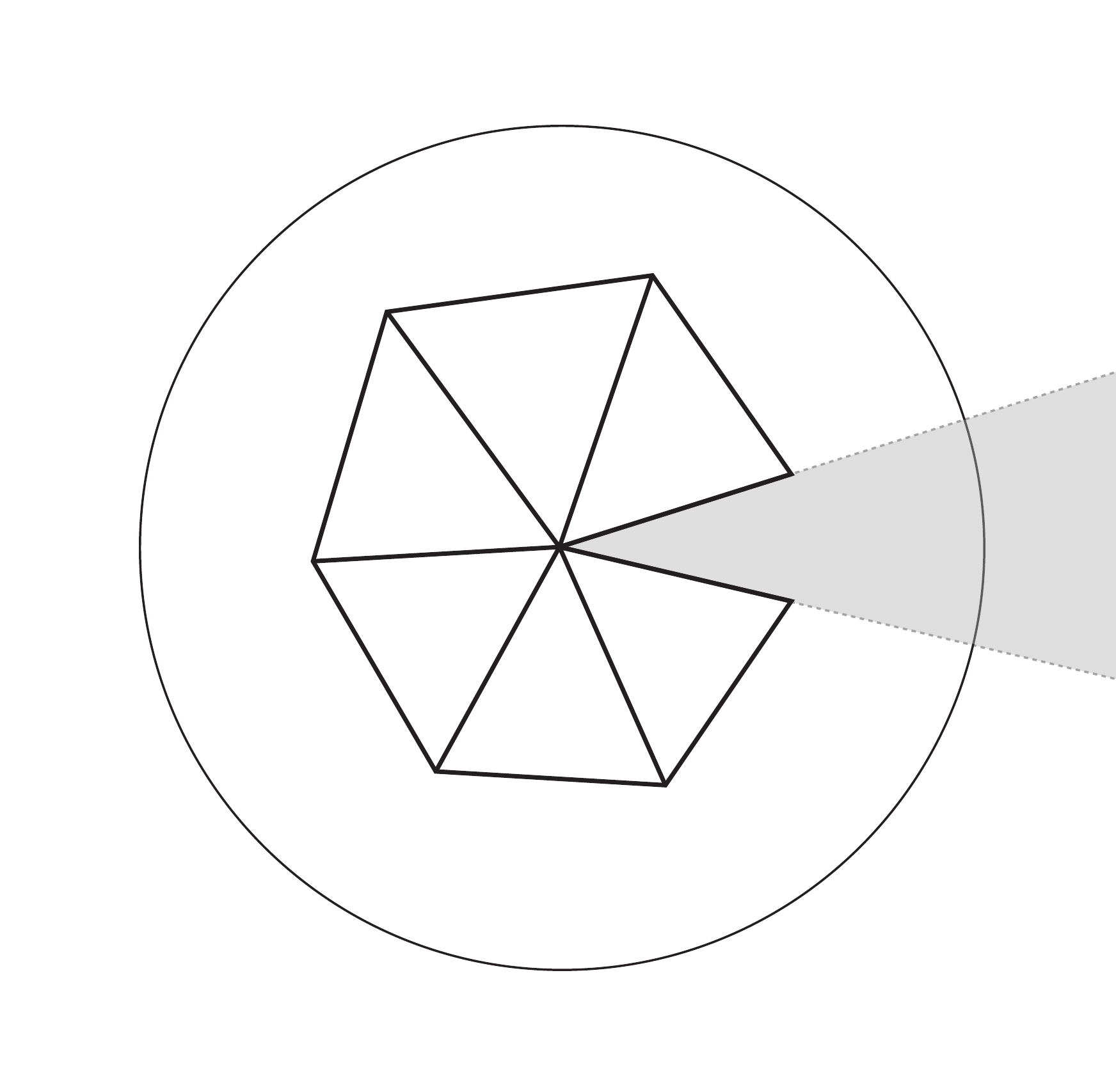}
\caption{\normalfont%
An example of a 2-dimensional lattice (with just one bone) cut open and laid flat in a
Euclidean space. The shaded wedge is not part of the lattice and its radial edges should be
identified. The defect for this bone is exactly equal to the angle subtended by the wedge. The
unit circle is drawn simply to remind us that the signature is Euclidean.}
\label{fig:fig-03}
\end{figure}

\begin{figure}[t]
\Figure{0.75}{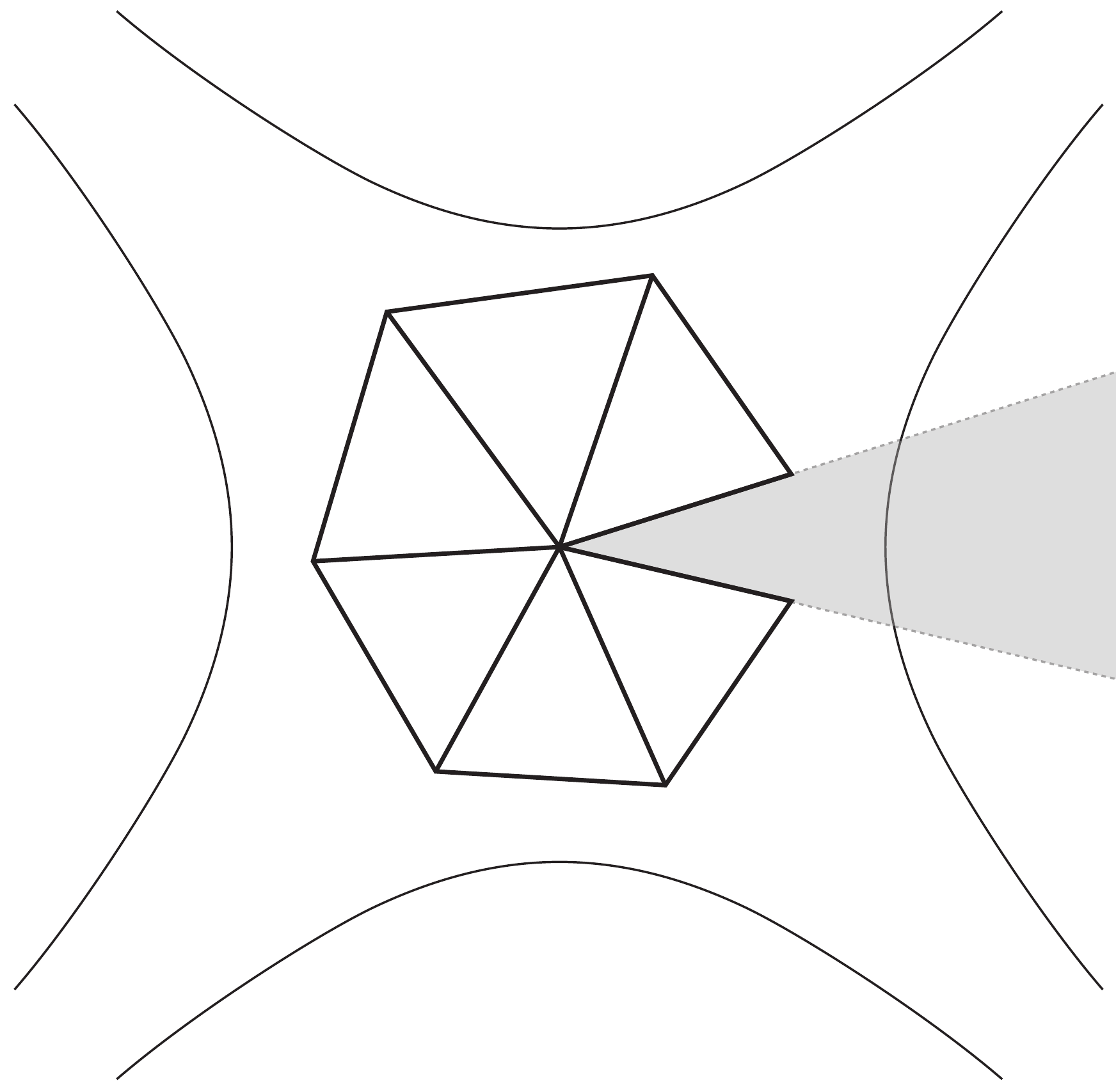}
\caption{\normalfont%
A similar construction to figure (\ref{fig:fig-03}) but in this instance for a Lorentzian
signature. Notice the four branches of the unit hyperbola, these match those from figure
(\ref{fig:fig-01}). In this example the defect is positive.}
\label{fig:fig-04}
\end{figure}

\begin{figure}[t]
\Figure{0.75}{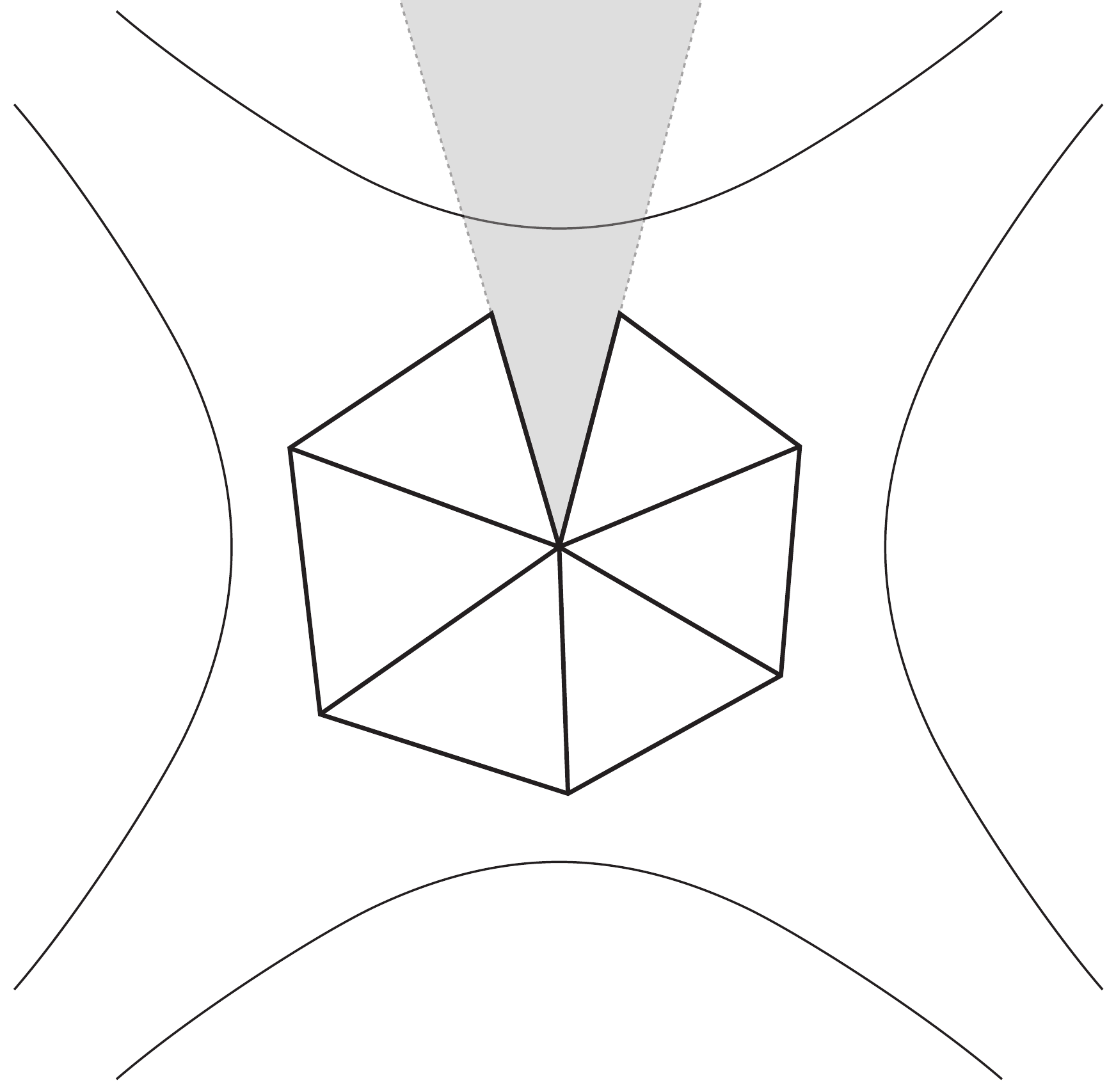}
\caption{\normalfont%
Here we have a case were the triangles were cut along a timelike edge. In this case we have a
negative defect angle (and thus this is not the same lattice as shown in the previous figure).
}
\label{fig:fig-05}
\end{figure}

\begin{figure}[t]
\Figure{0.75}{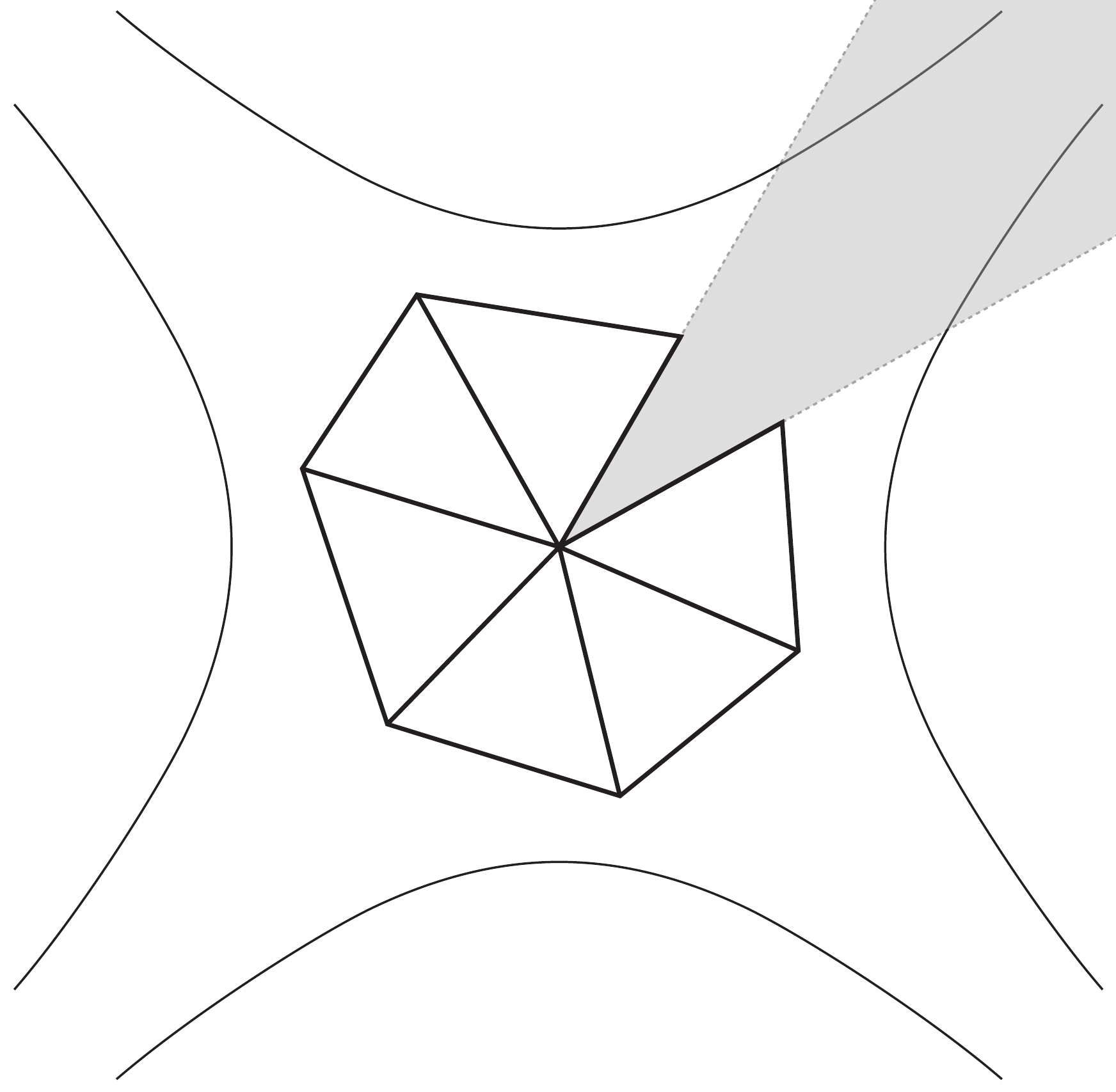}
\caption{\normalfont%
This arrangement is impossible. The two radial edges of the wedge have different signatures
and thus they can not be identified.}
\label{fig:fig-06}
\end{figure}

\begin{figure}[t]
\Figure{0.95}{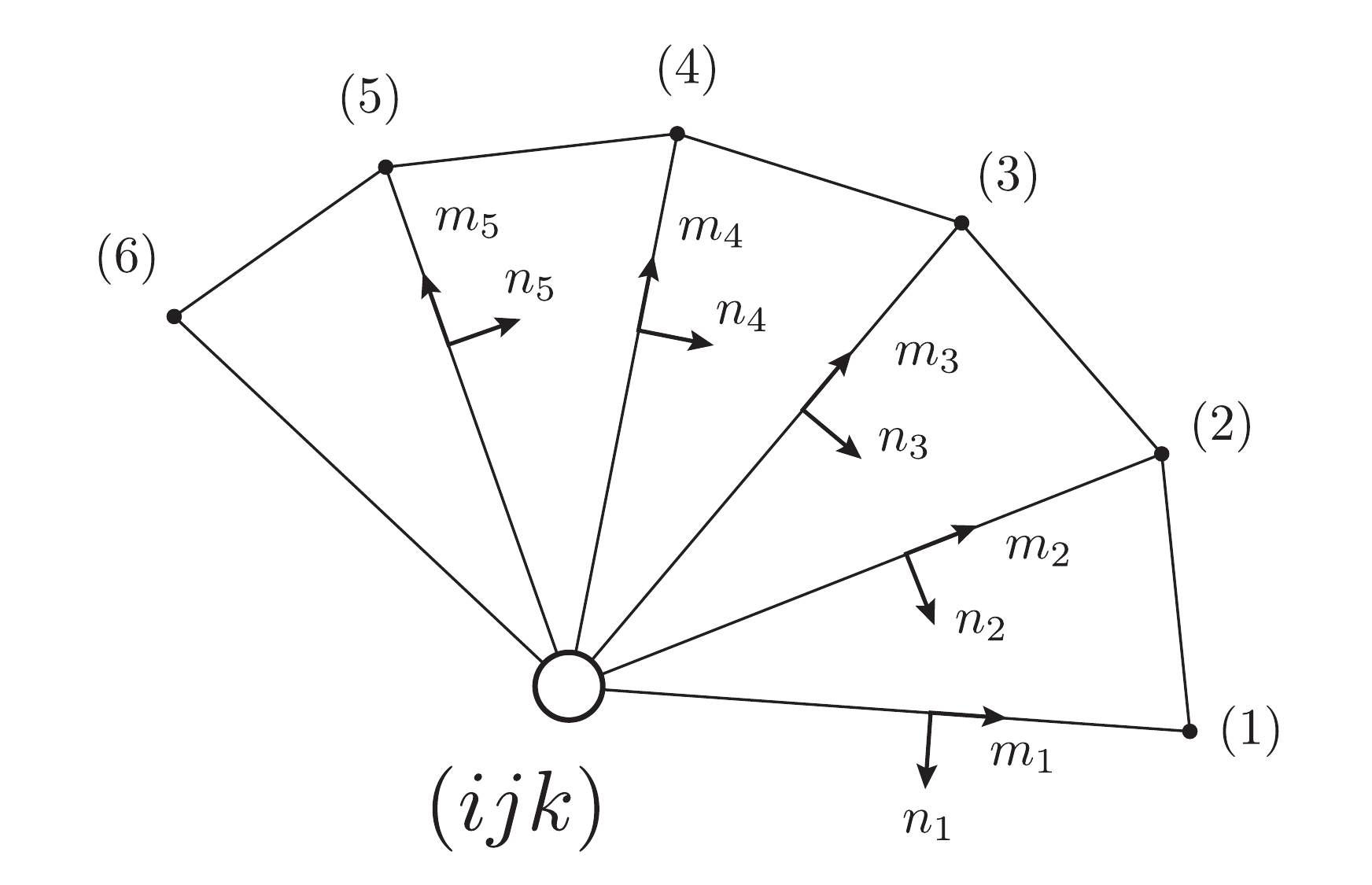}
\caption{\normalfont%
Here we show the first few 4-simplices attached to the bone $(ijk)$. Each pair of vectors here
are the basis vectors for the entry face for each 4-simplex. This figure is \emph{not} a map
of the simplices into a single flat space in the manner used in figures
(\ref{fig:fig-04}--\ref{fig:fig-06}). It is just a convenient picture of the vectors and their
relationships to their simplices.}
\label{fig:fig-07}
\end{figure}

\begin{figure}[t]
\Figure{0.85}{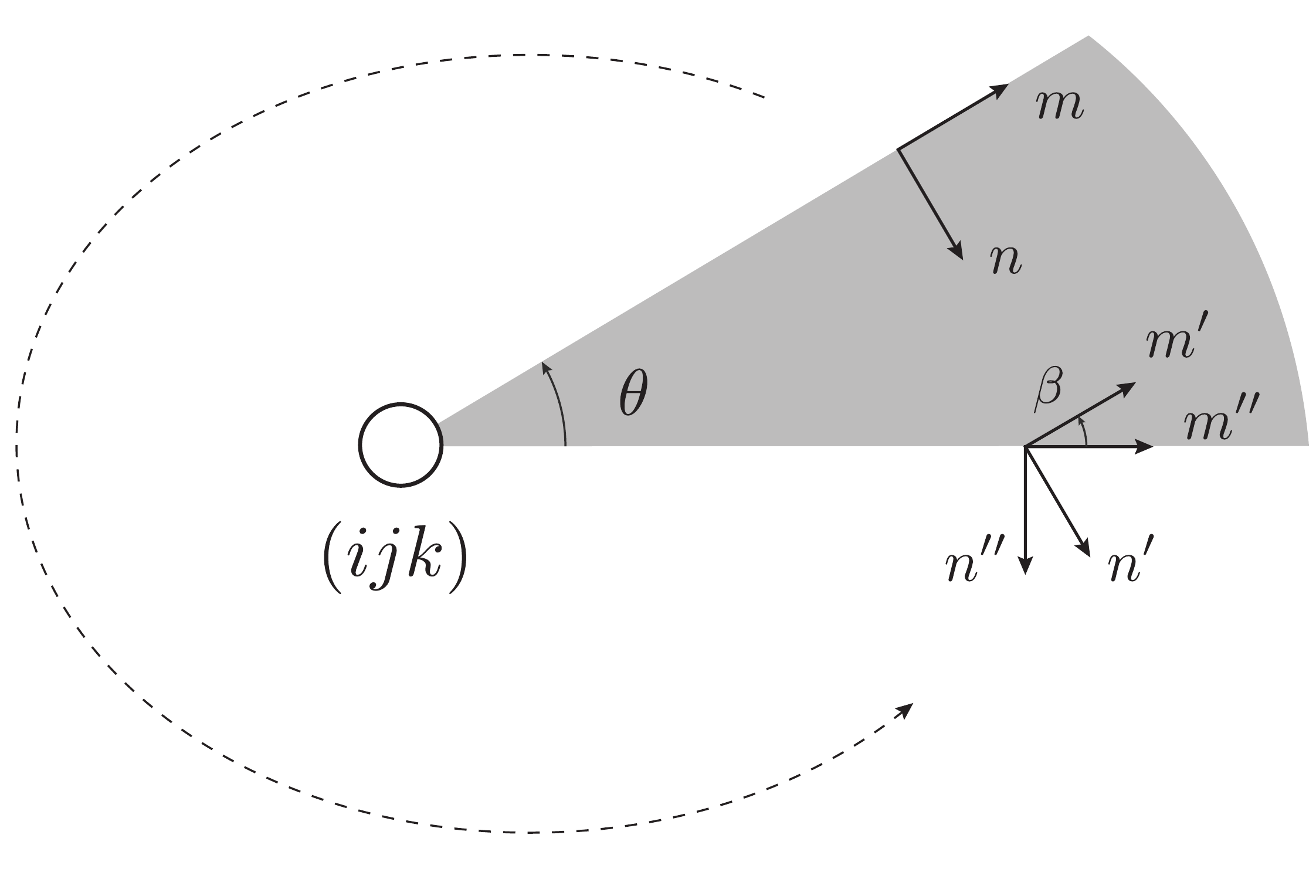}
\caption{\normalfont%
Here we show how the vectors $n$ and $m$ may be parallel transported around a bone. In
contrast to figure (\ref{fig:fig-07}) here we \emph{do} map the 4-simplices into a single
flat space, hence the introduction of the shaded wedge. This example is for a timelike bone.
By inspection we see that $\theta = \beta$.}
\label{fig:fig-08}
\end{figure}

\begin{figure}[t]
\Figure{0.85}{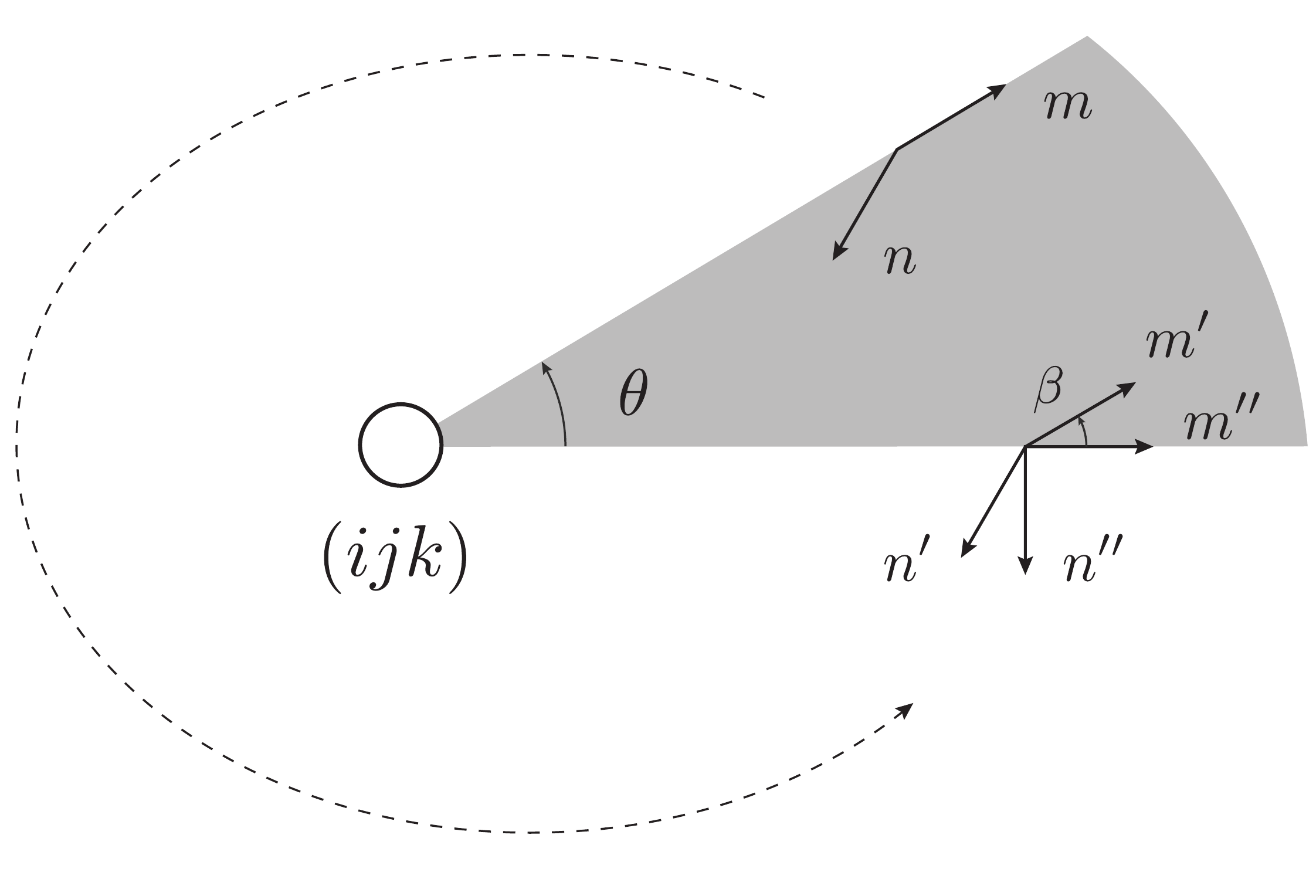}
\caption{\normalfont%
This is similar to the previous figure but modified for the case of a spacelike bone. We are
now working with boosts rather than rotations. Here we assume that $m$ is spacelike and thus
we must have $\beta<0$. We also see that $\theta>0$ thus we have $\theta=-\beta$. This is the
situation we expect if we do a parallel transport starting from $(ijk12)$ (see figure
\ref{fig:fig-07})}.
\label{fig:fig-09}
\end{figure}

\begin{figure}[t]
\Figure{0.85}{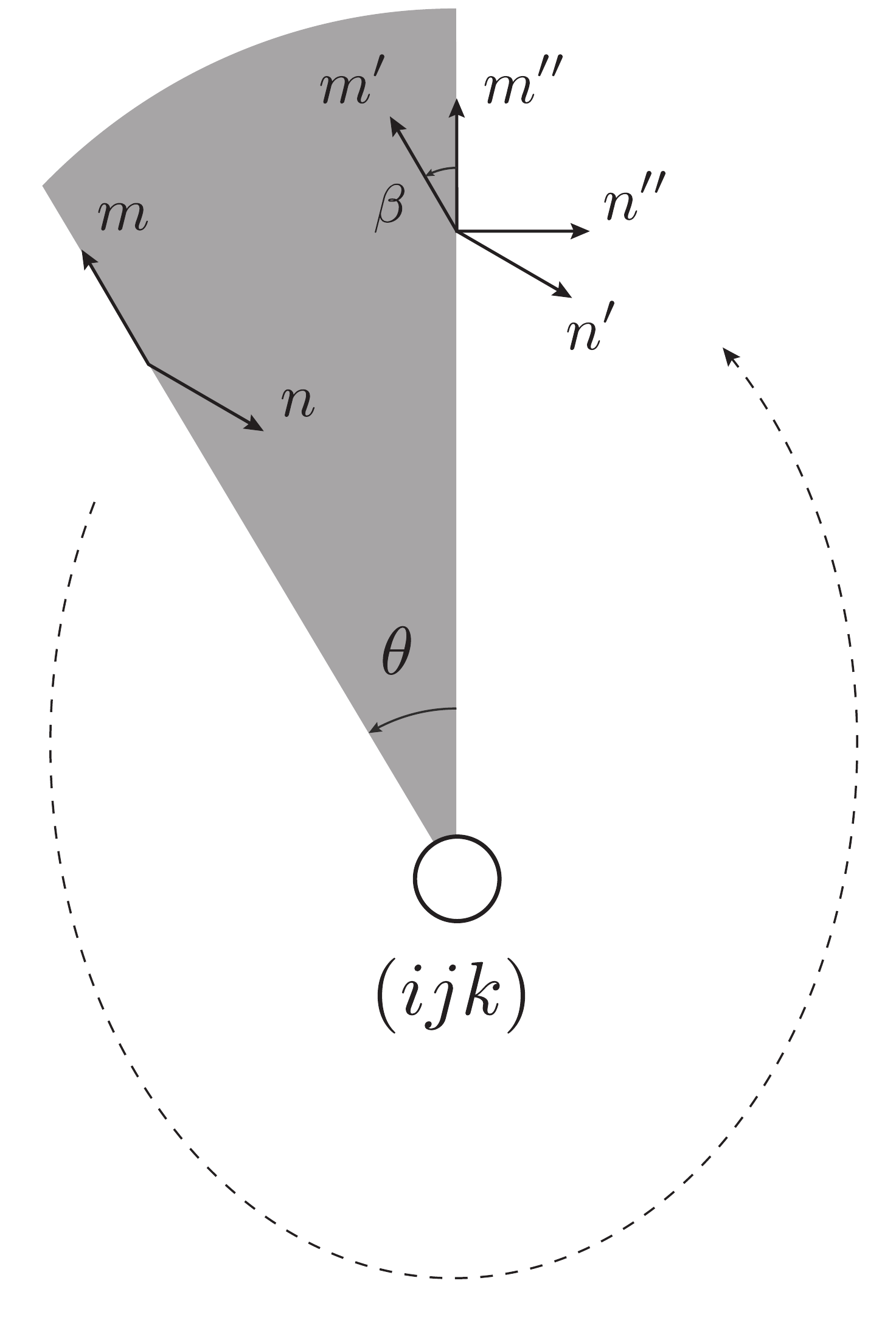}
\caption{\normalfont%
This is also for a spacelike bone but now our parallel transport loop starts and finishes
in $(ijk56)$. Here we see that $\beta<0$ and now $\theta<0$ (again, this is not the same
lattice as used in the previous figure). Now we find $\theta = \beta$.}
\label{fig:fig-10}
\end{figure}

\clearpage


\providecommand{\href}[2]{#2}\begingroup\raggedright\endgroup

\end{document}